\documentclass[journal,comsoc]{IEEEtran}

\usepackage{cite}

\ifCLASSINFOpdf
 
\else
 
\fi

\usepackage{caption}
\usepackage{algorithm}
\usepackage{algorithmic}
\usepackage{graphicx}
\usepackage{enumerate}
\usepackage{amsthm}
\usepackage{epstopdf}
\usepackage{amsmath,amssymb,epsfig,graphicx,slidesec,colortbl}
\usepackage[absolute,overlay]{textpos}
\usepackage{multimedia,multirow}
\usepackage{cite}
\usepackage{makecell}
\usepackage{subfig}
\usepackage{color}
\usepackage{url}
\usepackage{CJK}

\newtheorem{Def}{Definition}
\newtheorem{The}{Theorem}

\newtheorem{Cor}{Corollary}

\begin{document}

\title{Prospect Theoretic Analysis of  Privacy-Preserving Mechanism}

\author{Guocheng Liao,~\IEEEmembership{Student Member,~IEEE,}
        Xu Chen,~\IEEEmembership{Member,~IEEE,}
        and Jianwei Huang,~\IEEEmembership{Fellow,~IEEE}
    \thanks{This work was supported in part by the General Research Fund CUHK 14219016 from Hong Kong UGC, Presidential Fund from the Chinese University of Hong Kong, Shenzhen, and the Shenzhen Institute of Artificial Intelligence and Robotics for Society, and in part by the National Science Foundation of China (No. U1711265, No. 61972432); the Program for Guangdong Introducing Innovative and Enterpreneurial Teams (No.2017ZT07X355);the Pearl River Talent Recruitment Program (No.2017GC010465). Part of this paper was presented in GLOBECOM 2017 \cite{PTPrivacy}. 
    	
    	Guocheng Liao is with Department of Information Engineering, the Chinese University of Hong Kong (e-mail:
    		lg016@ie.cuhk.edu.hk). 
    		
    		Jianwei Huang is with the School of Science and Engineering, The Chinese University of Hong Kong, Shenzhen, China, the Shenzhen Institute of Artificial Intelligence and Robotics for Society (AIRS), and the Department of Information Engineering, The Chinese University of Hong Kong, Hong Kong, China (e-mail: jianweihuang@cuhk.edu.cn). 
    		
    		Xu Chen is with School of Data and
    		Computer Science, Sun Yat-sen University, Guangzhou, China (e-mail:
    		chenxu35@mail.sysu.edu.cn.)}}

\maketitle

\begin{abstract}

We study a problem of privacy-preserving mechanism design. A data collector wants to obtain data from individuals to perform some computations. To relieve the privacy threat to the contributors, the data collector adopts a privacy-preserving mechanism by adding random noise to the computation result, at the cost of reduced accuracy. Individuals decide whether to contribute data when faced with the privacy issue. Due to the intrinsic uncertainty in privacy protection, we model individuals' privacy-related decision using Prospect Theory. Such a theory more accurately models individuals' behavior under uncertainty than the traditional expected utility theory, whose prediction always deviates from practical human behavior. We show that the data collector's utility maximization problem involves a polynomial of high and fractional order, the root of which is difficult to compute analytically. We get around this issue by considering a large population approximation, and obtain a closed-form solution that well approximates the precise solution. We discover that the data collector who considers the more realistic Prospect Theory based individual decision modeling would adopt a more conservative privacy-preserving mechanism, compared with the case based on the expected utility theory modeling. We also study the impact of Prospect Theory parameters, and concludes that more loss-averse or risk-seeking individuals  will trigger a more conservative mechanism. When individuals have different Prospect Theory parameters, simulations demonstrate that  the privacy protection first becomes stronger and then becomes weaker  as the heterogeneity increases from a low value to a high one.

\end{abstract}


\IEEEpeerreviewmaketitle

\section{Introduction}
\subsection{Background and Motivation}

Personal data collection is becoming increasingly common in  our daily life, in various industries such as online social network and medical treatment, to better understand the individuals and gain new inspirations for knowledge generation. 

A privacy-aware individual, however, may have the privacy concern when being requested for data contribution. He worries about potential personal information leakage if the data collector does not provide enough privacy protection. A recent example of this is the data leakage incident of Facebook due to the illegal data usage of third party Cambridge Analytica \cite{facebook_cambridge}. He would carefully evaluate the protection consequence promised by the data collector.

The growing privacy concern brings great challenges to the data collector regarding the data analysis. She\footnote{In this paper, for the ease of presentation clarity, we will use ``he'' to refer to an individual and use ``she'' to refer to the data collector. Such a terminology choice does not reflect any gender bias.} conducts some  computations   with collected data (e.g., exploiting individuals' smoking habit to study its relation with the chance of getting the lung cancer).  She would adopt a privacy-preserving mechanism, which adds some random noise to the computation result such that an adversary cannot easily infer participants' actual information. The downside is that the added noise will reduce the accuracy of the result. For example, an inaccurate average number of cigarettes that people smoke per day may not fully indicate the accurate relation between the smoking habit and lung cancer. The collector needs to carefully design the privacy-preserving mechanism to trade off the individual satisfaction and the computation accuracy.

One key feature of this privacy-preserving data collection problem is the uncertainty of privacy protection level, as individuals are not always sure about the privacy level promised by the data collector. As suggested  in \cite{behavioral_privacy,rationality},  the uncertainty of outcomes plays a significant role in privacy-related decision making, and  behavioral economics can help better understand the individual's decision concerning privacy. When dealing with uncertainty, most prior studies applied the Expected Utility Theory (EUT) to model an individual's decision  that maximizes his expected utility. However, experimental evidences (e.g.,\cite{prospect1979}) showed that in practice human behavior could significantly deviate from the EUT, due to the complex psychological perception and subjectivity \cite{weightingfunction}. This indicates that the traditional EUT is not accurate enough to capture an individual's decision pattern.

Alternatively, Prospect Theory (PT) \cite{prospect1979,advances_prospect1992}, a Nobel-Prize-Winning theory in behavioral economics, can provide a more accurate prediction on an individual's behavior. Supported by a large number real-word experiments (e.g., \cite{Vietnamsurvey,usstyle}), Prospect Theory both normatively and descriptively interprets how individuals make decisions by evaluating uncertain gains and losses. This motivates us to consider Prospect Theory to model individuals’ privacy-related decisions under uncertainty. Furthermore, Prospect Theory has been successfully applied in many areas, such as cognitive radio network management (e.g.,\cite{PTinterfere,prospect_pricing_cognitive}) and smart grid management (e.g., \cite{PTenergystorage,prospect_microgrids}).  However, there does not exist any theoretical studies regarding the application of Prospect Theory in the area of privacy-preserving mechanism design. Our paper presents the first step towards this important and under-explored area.

In this paper, with prospect theoretic modeling,  we will explore the answers to the following key questions: 
\begin{itemize}
	\item  \emph{From the individuals' perspective, how would they decide whether to participate in the data collection considering the privacy protection uncertainty? }

	\item \emph{From the data collector's perspective, how would she design the privacy-preserving mechanism considering the individuals' behavior?}

\end{itemize}

To answer the above questions, we model the interaction between the data collector and individual as a two-stage Stackelberg game. At the lower level, we use Prospect Theory to capture individuals' subjective decision-making under the privacy protection uncertainty. At the higher level, a better privacy protection by the data collector will attract more individuals, but the corresponding higher level of perturbation (due to the added noise)  will degrade the accuracy of the analysis. We compute the data collector's optimal strategy based on her prediction of individuals' participation decisions to various privacy protection levels.

\subsection{Key Contributions}

The main contributions of this paper are as follows.

\begin{itemize}
	\item \emph{Prospect Theory-based individual behavior model}: Due to the effectiveness uncertainty of the privacy-preserving mechanism, we model individuals' decisions based on Prospect Theory, which is more accurate compared with the widely adopted approach of EUT. In particular, we focus on understanding the impact of both the level of loss aversion and the shift of reference point.

	\item \emph{Analysis of the data collector's utility maximization problem}. Since this problem involves a polynomial of high and fractional order, it is difficult to obtain the analytical solution. We consider a large individual population approximation that allows us to compute a unique optimal solution that is close to the optimal solution in realistic settings.

	\item  \emph{Design insights based on the impact of prospect theoretic model}: We compare the results under the prospect theoretic model and that under the EUT model, and conclude that the data collector should adopt a more conservative privacy-preserving mechanism based on Prospect Theory. Regarding  the impact of  Prospect Theory parameters, we find that more loss-averse individuals lead to a more conservative privacy-preserving mechanism. However, the shift of reference point towards a more tolerant attitude does not always indicate a less conservative mechanism. Taking into account the heterogeneity of different individuals' Prospect Theory parameters,  we find that the privacy protection first becomes stronger and then becomes weaker  as the heterogeneity increases from a low value to a high one.

\end{itemize}

The rest of this paper is organized as follows. We first introduce the related work in Section \ref{Literaturereview}. In Section \ref{Pre}, we introduce the preliminaries  of differential privacy, which is the privacy metric that we use in this paper.  In Section \ref{SystemModel}, we discuss the system model regarding the individual's participation problem and the data collector's utility maximization problem, respectively. In Section \ref{Solving}, we solve these problems and analyze the impact of different Prospect Theory parameters. In Section \ref{Simulation} and Section \ref{sec_heterogeneity}, we provide simulation results and discuss the corresponding insights. We conclude the  paper in Section \ref{Conclusion}.

\section{Literature Review}\label{Literaturereview}

The problem of privacy-preserving data collection has been widely considered based on the concept of differential privacy\cite{calibrating}, which is regarded as a powerful tool to quantify privacy in the literature (e.g., \cite{privacy_endogenous,approximately_auction_cost_data,selling_privacy_auction,buyingwithoutverification,value_privacy,game_quality}). Differential privacy requires that an individual' data only has a limited impact on the output of a computation, therefore it is hard for an adversary to infer an individual's data from the exposed output.  Motivated by this, we use differential privacy as the privacy model in this paper. However, different from previous literature\cite{privacy_endogenous,approximately_auction_cost_data,buyingwithoutverification,value_privacy,game_quality,selling_privacy_auction},  we model the individuals' subjective reactions to the privacy metric with Prospect Theory due to the intrinsic uncertainty.

Regarding the problem of privacy-preserving data collection, the studies in \cite{privacy_endogenous,approximately_auction_cost_data,buyingwithoutverification,value_privacy,game_quality,selling_privacy_auction,nonmonetarygame,linearregressiongame} considered a  stylized case: the data collector,  with a certain computation goal, wants to obtain the personal data from privacy-aware individuals. However, the detailed problem model and analysis can be different based on the incentive methods adopted by different applications. There are mainly three categories: no incentive (e.g., \cite{privacy_endogenous}), monetary payments (e.g., \cite{selling_privacy_auction,approximately_auction_cost_data,buyingwithoutverification,value_privacy,game_quality,inception,bidguard,reap,bilateral}), and  non-monetary rewards (e.g., \cite{nonmonetarygame,linearregressiongame}). For the first category, Ghosh \emph{et al.} in \cite{privacy_endogenous} considered that an individual would participate if the privacy protection offered by the data collector satisfies his privacy requirement. For the second category of monetary payments, the studies in \cite{approximately_auction_cost_data,buyingwithoutverification,value_privacy,game_quality,selling_privacy_auction} focused on the mechanism design that minimizes the data collector's total payment subject to computation accuracy constraints. Crowd sensing (e.g., \cite{inception,bidguard,reap,bilateral}) is one business practice of this category. For example, Jin \emph{et al.} in \cite{inception} designed an auction-based incentive mechanism with data perturbation that ensures workers' privacy protection. Zhang \emph{et al.} in \cite{reap} designed a contract-based incentive mechanism that optimizes accuracy subject to monetary budget by satisfying different privacy preferences of different users.    For the third category, the studies in \cite{nonmonetarygame,linearregressiongame,meanfieldERM,socialcorrelation} considered that individuals can benefit from the data result computation, and focused on the individuals' trade-off between the privacy loss and the benefit. References \cite{meanfieldERM,socialcorrelation} further studied the privacy-preserving mechanism based on the individuals' trade-off.

Our model and analysis fall into the third category, since we do not rely on monetary payments but consider the analysis-based reward. In our problem, an individual needs to decide  whether to participate in the data collection.  In the studies in \cite{nonmonetarygame,linearregressiongame,meanfieldERM,socialcorrelation}, however, an individual decides on the variance of noise to be added to his reported data (as he will always participate).  Moreover, regarding the individuals' decision modeling, none of the previous studies (e.g., \cite{privacy_endogenous,approximately_auction_cost_data,buyingwithoutverification,value_privacy,game_quality,selling_privacy_auction,inception,bidguard,reap,bilateral,nonmonetarygame,linearregressiongame,meanfieldERM,socialcorrelation})  explored the  behavioral economics modeling of individuals.

Acquisti and Grossklags  in \cite{behavioral_privacy,rationality} suggested that behavioral economics is a powerful tool to better understand the privacy-related decision with empirical studies. However, they did not detail any behavioral theoretic model together with realistic privacy metric.  In our work, we fill in this gap based on a practical privacy-preserving framework, and  capture individuals' behavior under uncertainty by applying the Prospect Theory in behavioral economics.

 The adoption of Prospect Theory to better  design and analyze engineering systems only emerged recently (e.g., \cite{prospect_spectruminvestment,PTinterfere,prospect_pricing_cognitive,prospect_microgrids,PTcloud,PTjamminggames,PTUAV,PTHardware,PTEnergytrading,PTpower,PTcrowdsourcing,PTenergystorage,PTsatisfaction,PTdronedelivery,PTresilience,PTstochastic}). For example, Li \emph{et al.} in \cite{PTinterfere} studied a random access game in wireless networks, and compared the performance under Nash Equilibrium under for both Prospect Theory and EUT settings. Yang \emph{et al.} in \cite{prospect_pricing_cognitive} considered the impact of user decision-making on radio resource pricing, and  proposed prospect pricing to improve radio resource management. Xiao \emph{et al.} in  \cite{prospect_microgrids} formulated a static energy exchange game between microgrids, and derived Nash Equilibrium to analyze the selling and buying decisions based on Prospect Theory. Saad \emph{et al.} in \cite{PTHardware} considered the security of integrated circuit outsourcing, and captured the subjectivity of attacker and defender with weighting effect in Prospect Theory.  These previous studies (e.g., \cite{PTinterfere,prospect_microgrids,PTcloud,PTjamminggames,PTUAV,prospect_pricing_cognitive,PTHardware}) applied  one perspective of probability distortion in Prospect Theory. In our work, we consider two perspectives of S-shape valuation function and reference point to capture individuals' subjective perception of uncertainty. Due to the continuous nature of privacy outcome, we will leave the more complicated analysis of probability distortion in the future work. Furthermore, to the best our knowledge, our work is the first  theoretical research exploring the application of Prospect Theory in privacy-preserving mechanism design.

\section{Preliminaries}\label{Pre}
Our individuals' decision model is based on the Prospect Theory, and our privacy model is based on  the widely used theoretical framework of differential privacy with privacy guarantee. In this section, we introduce the preliminaries on Prospect Theory and differential privacy.

\subsection{Prospect Theory}
\begin{figure}[!t]
	\centering
	\includegraphics[width=3.3in]{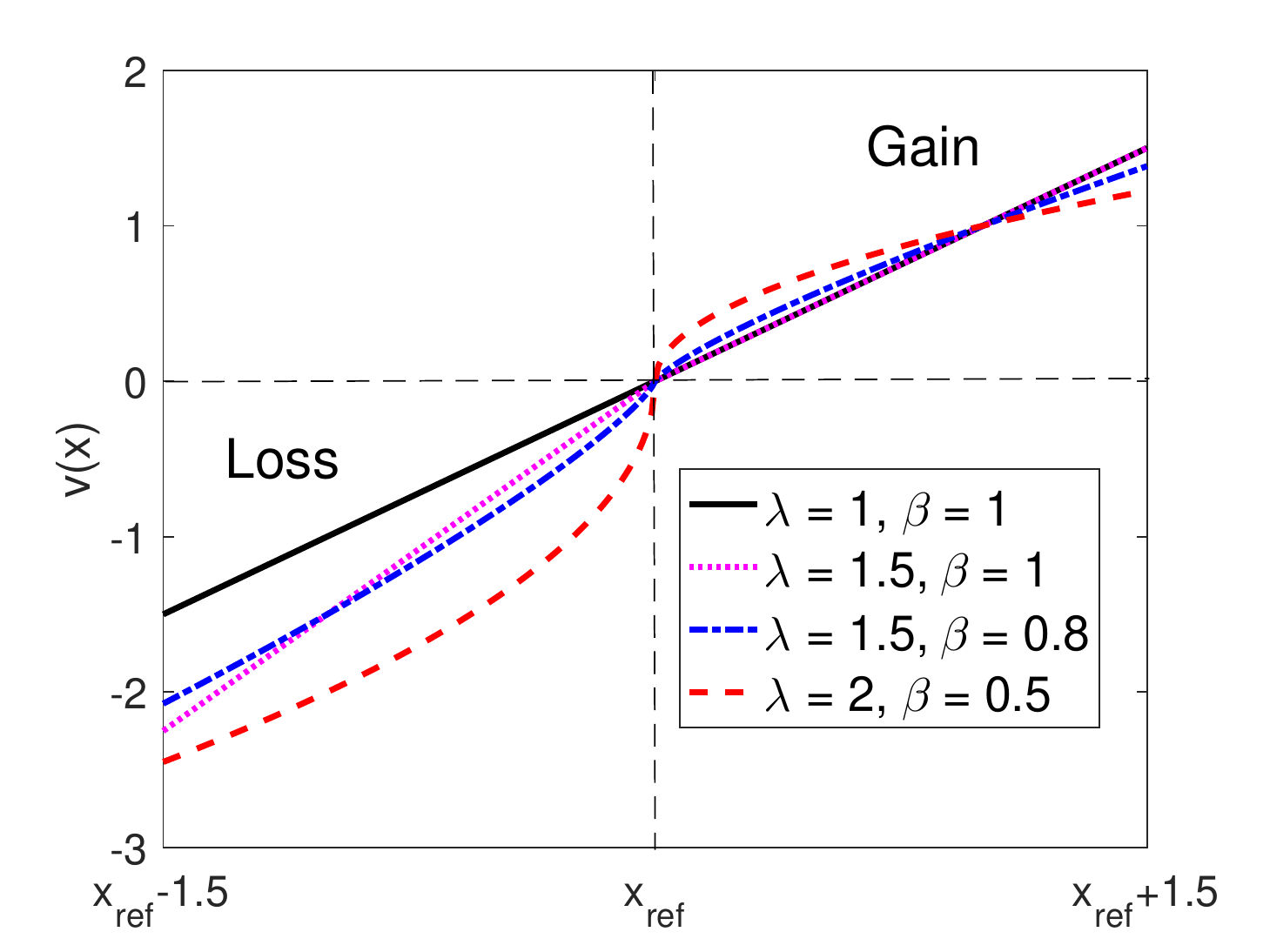}
	\caption{S-shape asymmetrical valuation function.}
	\label{valuationfunction_1}
\end{figure}

Compared with the widely-used modeling approach of expected utility theory in economics, Prospect Theory better captures practical human behavioral characteristics under uncertainty. Prospect Theory suggests that an individual evaluates an outcome with his subjective perception due to psychological loss and risk preference. An S-shape asymmetrical valuation function together with a reference point \cite{advances_prospect1992} (as shown in Fig \ref{valuationfunction_1}) characterizes such a pattern: 

\begin{equation}\label{valuation}
v(x) = \left\{ {\begin{array}{*{20}c}
	\begin{aligned}
	&(x-x_{ref})^\beta, &\text{if}\  x \ge x_{ref}\\ 
	&-\lambda(x_{ref}-x)^\beta, & \text{if}\  x<x_{ref}\\	
	\end{aligned}
	\end{array}}\right..
\end{equation}
Here $x$ represents the actual outcome, e.g., the amount of money earned in the gambling. A larger $x$ indicates a better outcome. Parameter $x_{ref}$ represents the reference point, and $\lambda \ge 1$ and $\beta\in (0,1]$ are loss aversion parameter and risk parameter, respectively. We will explain the physical meaning of these parameters later. The valuation $v(x)$ describes how individual subjectively evaluates the actual outcome. 

\subsubsection{Reference Points}
Reference point serves as the individual's benchmark to evaluate the actual outcome. If the actual outcome is higher than the reference point, the individual will perceive it as a gain, due to a better outcome than his anticipation (represented by his reference point). Otherwise, he will perceive it as a loss. For example, an gambler targets at earning 100 dollars, which is his reference point. Whenever he earns less than 100 dollars (even though he indeed earns some money), he would consider it as a loss. 

\subsubsection{Loss and Risk Parameters}
The loss penalty parameter $\lambda$ captures the loss aversion  level. As $\lambda$ is often greater than one, we know that the impact of the loss is larger than that of the gain of the same absolute value. For example, when an gambler losses 100 dollars, he feels as if he lost more than 100 dollars, due to his loss aversion attitude. A larger $\lambda$ indicates that he is more loss averse. The parameter $\beta$ describes the concavity of the gain part of the function and the convexity of the loss part of the function, capturing the risk aversion level toward the gain and the risk-seeking level toward the loss.  Finally, when $\beta=\lambda=1$, the S-shape function degenerates to a straight line, which corresponds to the Expected Utility Theory  \cite{EUT} in the literature.

Finally, the prospect outcome is obtained by associating the valuation function with probability weighting:

\begin{equation}\label{equ_pt}
U_{PT} = \sum_{i}v(x_i) \cdot p_i, 
\end{equation}
where $p_i$ is the probability weighting associated with the outcome $x_i$.\footnote{The general formula of a prospect outcome also integrates probability weighting distortion to $p_i$. In this work, however, we do not consider the weighting distortion for simplicity, since  the continuous and infinite nature of privacy outcome would complicate the associated weighting analysis.}

\subsection{Differential Privacy}
Definition \ref{definition_differential_privat} defines $\epsilon$-differential privacy, which serves as a basic framework to offer reliable privacy protection in the related literature \cite{selling_privacy_auction,approximately_auction_cost_data,buyingwithoutverification,value_privacy,game_quality}. In the definition, one entry in a database corresponds to one individual's reported data. A  differentially private mechanism ensures that the result of the computation will not change significantly when an individual's data is added to the database. This ensures that when the computation result is revealed, an adversary can not easily infer the information of an individual's data without extra information.  
\begin{Def}{($\epsilon$-\bf{differential privacy})\cite{calibrating}} \label{definition_differential_privat}
	A randomized mechanism $\mathcal{A}$ is $\epsilon$-differentially private if for any two neighboring databases $D,D'$ that only differ in only one entry and for any set $\mathcal{S}$ of outputs:
	\begin{center}
		$\text{Pr}[\mathcal{A}(D) \in \mathcal{S}] \le \exp (\epsilon) \cdot \text{Pr}[\mathcal{A}(D') \in \mathcal{S}] $,
	\end{center}
	where $Pr[\cdot]$ denotes the probability of the event.
\end{Def}

Consider the extreme case of $\epsilon=0$, i.e., $exp(\epsilon)=1$, Definition \ref{definition_differential_privat} implies that $\text{Pr}[\mathcal{A}(D) \in \mathcal{S}]=\text{Pr}[\mathcal{A}(D') \in \mathcal{S}] $. This means that any two neighboring databases will have the same output distribution regardless of any single entry difference, which means a perfect privacy protection. When the value $\epsilon$ becomes \emph{larger}, the privacy protection becomes \emph{weaker}. 

\subsection{Laplace Mechanism}

Laplace mechanism is a commonly used mechanism that can ensure differential privacy for numerical data\cite{calibrating,privacy_endogenous,selling_privacy_auction,approximately_auction_cost_data}. Such a mechanism adds random noise with a Laplace distribution to the computation result and calibrates the standard deviation of the noise according to the sensitivity of the computation function (defined as follows).

\begin{Def}{(\bf{Sensitivity}) \cite{calibrating}}
	The sensitivity of a function $h: \mathcal{D} \rightarrow \mathcal{R}$ is:
	$$ S(h) = \max\limits_{D,D'\in \mathcal{D}} ||h(D)-h(D')||,$$ for all neighboring databases $D$ and $D'$ that differ in only one entry.
\end{Def}
The sensitivity of a function measures the maximum variation that any single variable can cause to the computation result. For example, the mean function with the representation $h(X)=\sum\limits_i {x_i}/n$, where $X=[x_1,x_2,...,x_n]$ is the collected data,  has the sensitivity of $S(h)=x_{\max}/n$. Here, $x_{\max}=\max \limits_{i}|x_i|$. 
The frequency function (e.g., $h(X)=|\{x_i|x_i>0\}|/n$ that indicates the proportion of positive data in the whole dataset), has the sensitivity of $1/n$.

A Laplace distribution, denoted by $Lap(b)$ with the scale parameter $b$, has a probability density function: 
\begin{equation}\label{pdfLaplace}
p(x)=\frac{1}{2b} \exp(-\frac{|x|}{b}).
\end{equation}
It has a zero mean and a standard deviation of $\sqrt{2}b$. By calibrating the standard deviation according to the sensitivity of computation, the mechanism can achieve the differential privacy. 

\begin{The}{(\bf{Laplace mechanism})\cite{calibrating,algorithmicfoundation}}\label{Laplacemechanism}
	For any function $h: \mathcal{D} \rightarrow \mathcal{R}$, the following Laplace mechanism can achieve $\epsilon$-differential privacy:$$\mathcal{A}(D)=h(D)+Y,$$ where $Y$ is a random variable drawn from the Laplace distribution $Lap(S(h)/\epsilon)$ (see Section 3.3 in \cite{algorithmicfoundation} for a detailed proof). 
\end{The}

\section{System Model}\label{SystemModel}

Fig. \ref{Data collection} illustrates the system model of the privacy-preserving data collection process. In this model, a data collector wants to collect data from individuals, and provides an analysis-based reward as an incentive. First, the data collector designs a privacy-preserving mechanism. Second, individuals decide whether to report data  based on the trade-off between reward and privacy loss. Next, we describe the individual's participation problem and the data collector's utility maximization problem in Section \ref{sec_Ind_par} and \ref{Sec_Collector_utility}, respectively.

\subsection{An individual's Participation Problem}\label{sec_Ind_par}

In this subsection, we formulate individuals' participation problem under the privacy protection uncertainty. We use  Prospect Theory to model individuals' behavioral characteristics in this context. 
\subsubsection{Privacy Measurement Based on Differential Privacy}

Based on Definition \ref{definition_differential_privat} of the $\epsilon$-differential privacy, we regard $\epsilon$ as the \emph{privacy level} for a given mechanism. Recall that a larger value of $\epsilon$ means a weaker privacy protection, and a smaller value of $\epsilon$ means stronger privacy protection.

\begin{figure}[!t]
	\centering
	\includegraphics[width=3.3in]{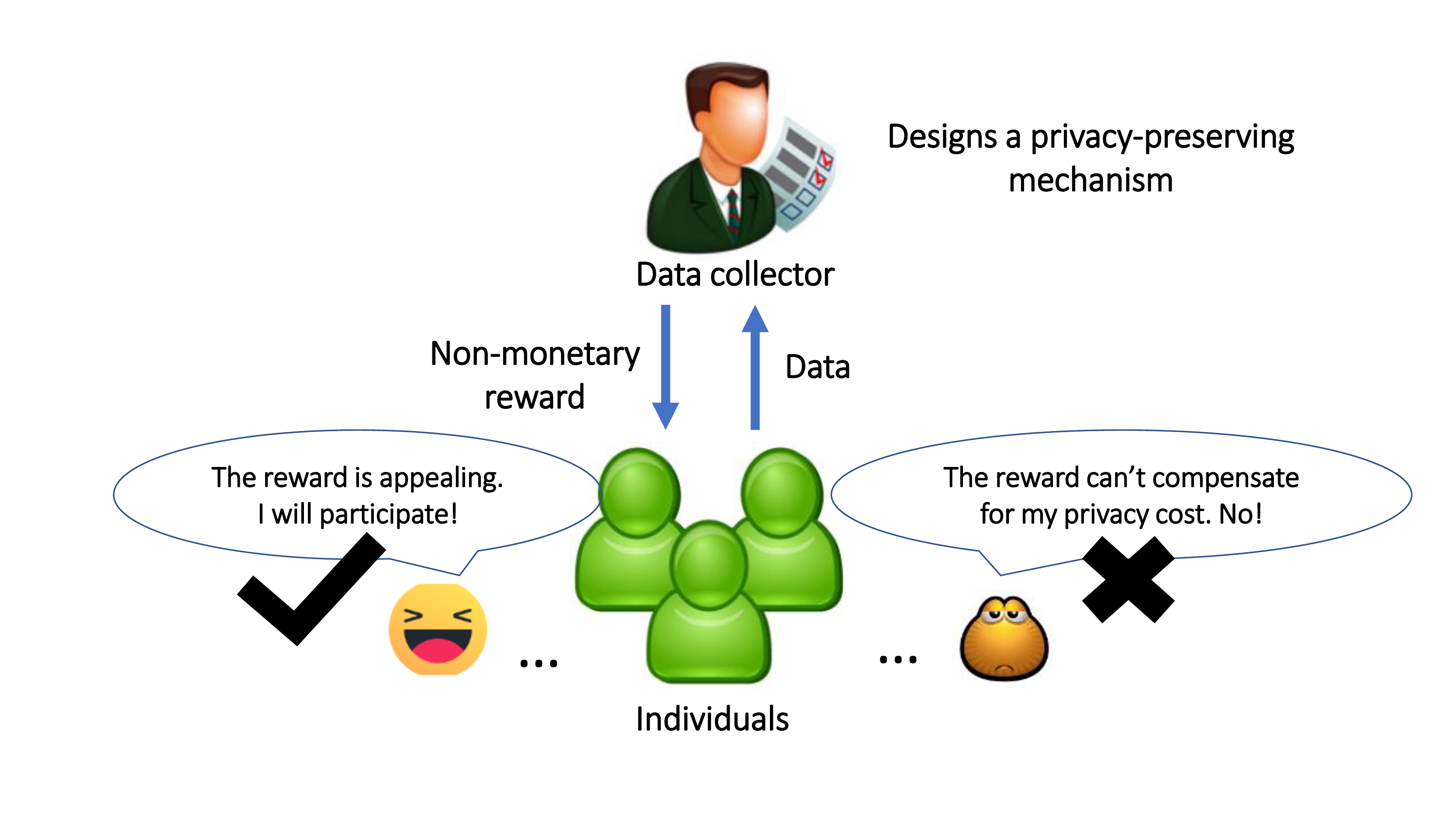}
	\caption{System model: in Stage \uppercase\expandafter{\romannumeral1}, the data collector initiates a collection with a privacy-preserving mechanism. In Stage \uppercase\expandafter{\romannumeral2}, individuals decide whether to participate in the data collection.  }
	\label{Data collection}
\end{figure}

As we can see from  Definition \ref{definition_differential_privat}, the privacy level involves some inherent uncertainty. The parameter $\epsilon$ corresponds to the weakest privacy protection (or the highest privacy leakage) among all possible neighboring databases. More specifically, data from different participants (i.e., different entries) may have different effects on the output, and this definition measures the most significant effect among all possible single entries. So the actual privacy level of a particular participant under this mechanism can be uncertain. This motivates us to use the Prospect Theory to model how a participant subjectively responds when he concerns about the potential risk of his actual privacy level.

\subsubsection{Prospect Theoretic Model of an Individual's Preference}

We first characterize individual's subjective valuation on a particular actual privacy level. By applying the S-shape asymmetrical valuation function in (\ref{valuation}), we obtain the valuation function in our context:
\begin{equation}\label{pro_valuation}
v(\epsilon) = \left\{ {\begin{array}{*{20}c}
	\begin{aligned}
	&(\epsilon_{ref}-\epsilon)^\beta, &\text{if}\  \epsilon \le \epsilon_{ref}\\ 
	&-\lambda(\epsilon-\epsilon_{ref})^\beta, & \text{if}\  \epsilon>\epsilon_{ref}\\	
	\end{aligned}
	\end{array}}\right.,
\end{equation}
where $0<\beta\le1$, $\lambda\ge1$, and $\epsilon_{ref}$ is the reference point. One main difference is that in (\ref{valuation}), a larger value of $x$ corresponds to a better outcome, while in the differential privacy setting, a smaller value of $\epsilon$ corresponds to a better outcome of protection. So if the privacy level $\epsilon$ is lower than the reference point $\epsilon_{ref}$, the individual would consider treat it as a gain. Otherwise, he would treat it as a loss.

Next, we characterize the  prospect privacy level given an $\epsilon$-differential private mechanism. A challenge of applying the classical Prospect Theory in our setting comes from the fact that we consider the continuous privacy level (while most prospect theoretic analysis considered discrete outcomes, e.g.,  \cite{PTinterfere,prospect_pricing_cognitive,prospect_spectruminvestment,prospect_microgrids,PTcloud,PTjamminggames,PTUAV}). We adopt the result from \cite{continuou_sprospect_approximation} to get around this issue, by approximating the infinite number of continuous outcomes with a finite number of discrete outcomes. More specifically, we decompose the set of all possible continuous outcomes $[0,\epsilon]$ into $m$ discrete outcomes $i\epsilon/m, \  i$$=$$1,2,...,m$. Then a participant's prospect privacy level is the summation of weighted valuations of all discrete outcomes $p_iv(i\epsilon/m),\  i$$=$$1,2,...,m$, where $p_i$ is the weighting (or probability) assigned to the corresponding outcome. For computational simplicity, we assume that the probability is evenly assigned to all the outcomes,\footnote{By adopting the same method in \cite{continuou_sprospect_approximation}, the calculation can be extended to other distributions.} then we obtain the prospect privacy level under an $\epsilon$-differentially private mechanism as in (\ref{equ_pt}):
\begin{equation}\label{prospect_level2}
\epsilon_p = \frac{1}{m}\sum\limits_{i=1}^{m}v\left(\frac{i}{m}\epsilon\right).
\end{equation}

\subsubsection{Individuals' Utility Maximization Problem} We will derive the individuals' utility  maximization problem by discussing the privacy cost and the reward gain. We first introduce the participants' privacy cost and participation benefit, and then introduce the non-participants' privacy cost. The individuals decide whether to participate after comparing two options.

\emph{Privacy Cost of a  Participant:} When a participant's data is used in an $\epsilon$-differentially private mechanism, he will experience a privacy cost that is associated with the privacy level.\footnote{We do not consider the privacy cost from membership or non-membership to avoid trivial case. We only consider the privacy cost from participation when his data is being used.} Similar to \cite{selling_privacy_auction, privacy_endogenous, approximately_auction_cost_data,reap}, we model this privacy cost as a linear function of differential privacy level. Let $g(\epsilon_p)$ denote the linear function that maps the prospect differential privacy level to privacy cost, i.e., $g(\epsilon_p) = c \cdot \epsilon_p$. Here the parameter $c$ measures the privacy cost per privacy level. Since the data collector gathers the same type of data for a computation (e.g., income or movie rating), we assume that participants experience the same cost parameter c (similar as in \cite{value_privacy,game_quality}). Regarding the prospect theoretic model parameters (i.e., $\epsilon_{ref}$, $\lambda$, and $\beta$), we will first assume that they are homogeneous across all individuals in the analysis. Later in Section \ref{sec_heterogeneity}, we will further numerically explore the impact of heterogeneous parameters.

\emph{Participation Benefit:} We assume that participants can benefit from the non-monetary reward (e.g., computation analysis-based reward and other formats of benefit) from the data collector. For example, the participants contribute ratings on movies and obtain movie recommendations in return. The consumers report the experience of a product surveyed by a company, which could help improve the product and provide better service. The individuals' benefit from data contribution (e.g., their interest in uncovering new results) could be utilized to incentivize their participations, which helps the data collector get rid of further monetary cost. The consideration of  non-monetary reward has not only been theoretically considered and analyzed in  \cite{nonmonetarygame,linearregressiongame}, but also has been implemented in practical businesses \cite{alibabasurvey}. To characterize the impact of the benefit on individuals' decisions, we consider  \emph{an individual $i$'s valuation on the benefit} denoted by $W_i$ (measured in the same unit of the privacy cost). Different individuals might value the benefit differently.

\emph{Privacy Cost of a  Non-Participant:} For a non-participating individual, he will not get a reward from the data collector. Furthermore, his actual privacy level is zero, i.e., perfect privacy protection. Hence his utility is $g(v(0))=g(\epsilon^\beta_{ref})$. If the reference point $\epsilon_{ref}$ is positive,\footnote{A positive reference point could correspond to the case where the individual believes that participating in the data collection is the social norm.} then not participating indicates a ``gain" of privacy. If the reference point $\epsilon_{ref}$ is zero, then the utility would be zero, meaning no gain or loss of privacy.

\emph{Utility Maximization Problem:} The utility of a participant is the summation of benefit valuation and privacy cost, i.e., $W_i+g(\epsilon_p)$  where the cost takes a negative value. The utility of a non-participant is just his privacy cost. Each individual $i$ decides whether to participate in the data collection by solving the following optimization problem:

\begin{equation}\label{individual_participation}
\begin{aligned}
\max_{a_i} \ & \ U_i(a_i)=a_i(W_i+g(\epsilon_p))+(1-a_i)g(v(0)) \\
\text{s.t.} \ & \ a_i \in\{0,1\}.
\end{aligned}
\end{equation}
Action $a_i=1$ means participation, and $a_i=0$ otherwise. Similar to \cite{selling_privacy_auction,approximately_auction_cost_data}, we assume the data collector is trustworthy, hence all participants will truthfully report their data due to the trusted data collector. \footnote{The mechanism design problem involving untruthful data reporting due to a potentially untrustworthy data collector would be much more complicated \cite{value_privacy,game_quality,buyingwithoutverification}, and we will study it in the future work.}

\subsection{Data Collector's Utility Maximization Problem}\label{Sec_Collector_utility}

In this subsection, we discuss the data collector's computation and utility function. 

The data collector obtains data from individuals to perform a data-driven computation. Throughout this paper, we consider the case where the data collector wants to calculate the mean of the collected numerical data.\footnote{Our general prospect theoretic privacy-preserving mechanism  works for any other scenarios or applications involving privacy protection uncertainty.} This kind of analysis has a wide range of application scenarios. For example, the data collector wants to investigate the average salary level of residents in a district, or obtain the popularity of a new movie by investigating the average score rated by audiences. 

Then we discuss the components of the data collector's utility. She benefits more if she manages to collect more data, as it enables more convincing computation result \cite{useful_things_ml}. Meanwhile,  she adopts a differentially private mechanism  (e.g., the Laplace mechanism) that adds some random noise to the computation result, which leads to an accuracy penalty. This implies that the data collector's utility function depends on two factors: the amount of collected data and the accuracy penalty.  Recall that the data collector would provide non-monetary reward to incentivize individuals. The reward naturally comes from the data-driven analysis without incurring a significant additional cost, as such an incentive is a by-product of the data analysis.  The data collector does not have an additional cost (in terms of the incentives to individuals).

\subsubsection{Data Collector's Benefit}
\begin{figure}[!htbp]
	\centering
	\includegraphics[width=2in]{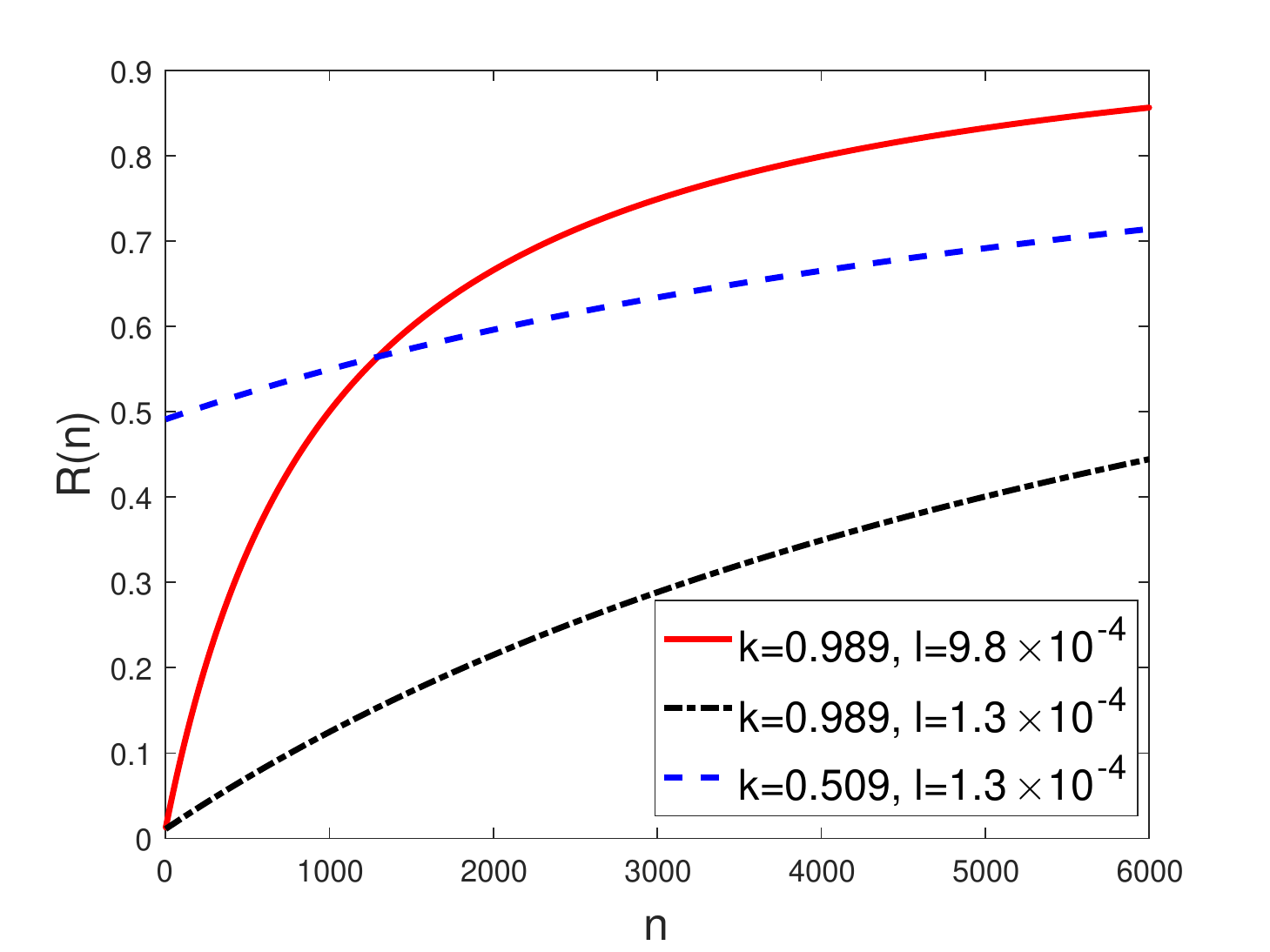}
	\caption{Data amount benefit function $R(n)$.}
	\label{dataamount}
\end{figure}

We use $R(n)$ to denote the data collector's benefit of collecting data from $n$ participants. We assume that $R(n)$ is non-negative, monotonic increasing, strictly concave, and upper bounded. As $n$ grows large, the marginal benefit of collecting data from one more participant reduces, hence the concavity shape of the function. Furthermore, the data amount is not the only factor that affects the computation result. Other factors such as methods of representation \cite{useful_things_ml} and optimization all influence the computation to some extent. Hence function $R(n)$ would be bounded even when $n$ goes to infinity. For the ease of expose, we follow \cite{IOT} and use the following benefit function with the parameters $k$ and $l$ in our analysis,
\begin{equation}\label{data amount component}
R(n,[k,l])=1-\frac{k}{1+l \cdot n}. 
\end{equation}
Here $k>0$ and $l>0$. Fig. \ref{dataamount} shows some examples of the function with different values of $k$ and $l$. The data collector needs to adjust the values of $k$ and $l$ to match the exact benefit of a particular application.

\subsubsection {Data Collector's Accuracy Penalty}

Define $l(x)$ as the penalty if the added noise level is $x$. The penalty is more significant if the noise multitude $|x|$ is larger, so $l(x)$ is non-negative and nondecreasing in $|x|$. Similar to \cite{utiltiy_max,universally_minimax}, we consider one of the possible representations, $l(x) = x^2$, which emphasizes the variance in the error. Meanwhile,  recall in Theorem \ref{Laplacemechanism} that the data collector can add Laplacian random noise to ensure $\epsilon$-differential privacy. Combining the probability density function in (\ref{pdfLaplace}), we get the expected accuracy penalty $L(\epsilon)$ under  $\epsilon$-differential privacy   as follows:

\begin{equation}\label{accuracy_penalty}
L(\epsilon)=\int_{-\infty}^{+\infty}l(x)p(x)dx=2\frac{S(h)^2}{\epsilon^2}. 
\end{equation}

The data collector needs to choose the privacy level $\epsilon$ to maximize her utility, i.e., 
\begin{equation}\label{collector_maximization}
\begin{aligned}
\max_{\epsilon >0} \ & \ U_c(\epsilon)=R(n(\epsilon))-L(\epsilon). \\
\end{aligned}
\end{equation}
Here $n(\epsilon)$ is the number of participants under the $\epsilon$-differentially private mechanism, which will be derived based on the individuals' responses to the mechanism (as in Section \ref{sec_Ind_par}).

\subsection{Problem Formulation}

We formulate the overall system as a two-stage game, as illustrated in Fig. \ref{Data collection}. In Stage \uppercase\expandafter{\romannumeral1}, the data collector designs an $\epsilon$-differentially private mechanism  to maximize its utility in (\ref{collector_maximization}). In Stage \uppercase\expandafter{\romannumeral2}, each individual decides whether to participate in the data collection  to maximize her utility in (\ref{individual_participation}). 
We use backward induction to solve this two-stage optimization problem.

\section{Solving the Two-stage Problem}\label{Solving}

\subsection{Individual's Decision-Making}
We first consider the case where individuals' reference point is zero, which means that individuals are intolerant and expect perfect privacy protection. Hence any privacy level induced in the data collection process will be considered as a loss. We will consider the case of a positive reference point later in Section \ref{sec_PTparameter}.

We will derive the number of participants under an $\epsilon$-differential private mechanism. We first analyze individuals' participation condition. In an individual's participation problem (\ref{individual_participation}), the individual will decide to participate if and only if $U_i(1)\geq U_i(0)$. Under the zero reference point, we have $g(v(0))=g(\epsilon_{ref}^\beta) = 0$, and the condition becomes  $W_i\geq g(v(0))-g(\epsilon_p)=-g(\epsilon_p)$.

We consider a group of $N$ individuals. Let $p_v(W)$ denote the probability density function of the  reward valuation $W$ among the individuals. Then the number of individuals choosing to participate in data collection is:
\begin{equation}\label{participants_number}
n(\epsilon)=N \cdot Pr\left(W>-g(\epsilon_p)\right)=N\int_{-g(\epsilon_p)}^{\infty}p_v(W)dW.
\end{equation}   
We can show that $n(\epsilon) $ is non-increasing with $\epsilon$. For the convenience of analysis, we let each individual's reward valuation $W$ follows a uniform distribution in $[0,W_{\max}]$  in the theoretical analysis in Section \ref{Solving}, i.e., 

\begin{equation}\label{reward_valuation_distribution}
p_v(W)=
\begin{cases}
\frac{1}{W_{\max}}, & \quad \text{if} \ W \in [0,W_{\max}]; \\
0, & \quad \text{otherwise}.\\
\end{cases}
\end{equation}

In Section  \ref{Simulation}, we will perform simulation studies based on a more general truncated normal distribution  (which includes the uniform distribution as a special case). 

\subsection{Data Collector's Optimal Differentially Private Mechanism}

Next we solve the data collector's utility maximization problem in (\ref{collector_maximization}). We assume that she possesses adequate information of the target individuals, i.e., the privacy loss coefficient $c$, the distribution $p_v(W)$ of reward valuation $W$, and Prospect Theory parameters, by abundant previous data-related investigations. Then she can decide the optimal $\epsilon$ to maximize her utility based on the anticipation of individuals' reactions.

Recall that we consider the mean analysis of numerical data as an example in this work. The range of the data is normalized from zero to one. The sensitivity of this computation is $S(h) = 1/n$. Then the accuracy penalty is: 
\begin{equation}\label{sensitivity}
L(\epsilon)=2\frac{S(h)^2}{\epsilon^2}=\frac{2}{n^2\epsilon^2}. 
\end{equation}

We substitute (\ref{participants_number}), (\ref{reward_valuation_distribution}) and (\ref{sensitivity})  to the data collector's utility maximization problem (\ref{collector_maximization}) and obtain a one-variable optimization problem. It is difficult to obtain the closed-form optimal solution, as the   derivative of the objective function involves a high-and-fractional-order polynomial. We can apply many effective one-dimensional search methods \cite{numerical_analysis} to numerically solve this problem. To gain useful insights for the practical implementation, nevertheless, we would like to derive an analytical solution by taking some reasonable approximations.

Next, we describe how  to derive the approximated optimal solution of (\ref{collector_maximization})  under a large population approximation. We first simplify the   derivative formulation and reduce the order by considering that the population size $N$ is large enough, and further get around the fractional order by considering that the parameter $\beta=1$. We then can compute the unique root of the   derivative in the feasible set, which is the approximated optimal solution.

More specifically, let us consider the feasible set of $\epsilon$ such that there is at least one participant (i.e, $n(\epsilon)>0$) according to (\ref{participants_number}) and (\ref{reward_valuation_distribution}):
\begin{equation}\label{feasibleset}
\left\{\epsilon:  -g(\epsilon_p) <W_{\max}\right\}.
\end{equation}

We can compute the data collector's objective function in (\ref{collector_maximization}) together with its   derivative as follows:

\begin{equation}\label{collector utility1}
U_c(\epsilon)=1-\frac{k}{1+l N\frac{W_{\max}+g(\epsilon_p)}{W_{\max}}}-\frac{2}{N^2\epsilon^2\left(\frac{W_{\max}+g(\epsilon_p)}{W_{\max}}\right)^2} ,
\end{equation}
and 
\begin{equation}\label{derivative}
U'_c(\epsilon) = \frac{klN\frac{g'(\epsilon_p)}{W_{\max}}}{\left[1+lN\frac{W_{\max}+g(\epsilon_p)}{W_{\max}}\right]^2}+ \frac{4}{N^2}\frac{\frac{W_{\max}+g(\epsilon_p)+g'(\epsilon_p)\epsilon}{W_{\max}}}{\left[\frac{W_{\max}+g(\epsilon_p)}{W_{\max}}\right]^3\epsilon^3} .
\end{equation}

One of the key challenges of analytically computing the root of $U'_c(\epsilon) = 0$ is due to the six-order polynomial at the numerator after combining the terms. However, we can approximate  $1+l N\left(W_{\max}+g(\epsilon_p)\right)/(W_{\max})$ with $l N\left(W_{\max}+g(\epsilon_p)\right)/(W_{\max})$ when the population size $N$ is large.  Hence we can obtain an approximated (denoted by the superscript $a$) version of (\ref{derivative}):
\begin{equation}\label{approx_derivative1}
U^{\prime a}_{c}(\epsilon)=\frac{4}{\left(\frac{W_{\max}+g(\epsilon_p)}{W_{\max}}\right)^3 \epsilon ^3 N^2}f(\epsilon),
\end{equation}
where

\begin{equation}\label{zero_poly}
f(\epsilon) = \left(\frac{W_{\max}+g(\epsilon_p)}{W_{\max}}\right)\left(1+ \frac{kN}{4l}\epsilon^{3}\frac{g'(\epsilon_p)}{W_{\max}}\right)+\epsilon\frac{g'(\epsilon_p)}{W_{\max}}.
\end{equation}
We can compute the root of $f(\epsilon)=0$ in the feasible as  the approximated optimal solution, which is unique.

\begin{The}\label{Theorem_optimalsolution}
	Under the approximation of (\ref{derivative}), we obtain the unique approximated optimal solution\footnote{We use notation $\tilde{}$ to represent an approximated optimal solution.} $\tilde{\epsilon}^*$ of the problem (\ref{collector_maximization}) in closed form as the root of $f(\epsilon)=0$.

\end{The}

We provide the proof of Theorem \ref{Theorem_optimalsolution} in Appendix A, including the detailed closed-form expression of the approximated optimal solution. The approximation enables us to derive a simplified polynomial function $f(\epsilon)$, which makes it mathematically tractable and straightforward to study the impact of Prospect Theory parameters on the optimal mechanism  solution (as we will do in Sections \ref{Sec_CMP_EUT} and \ref{sec_PTparameter}). The polynomial function serves as a bridge connecting Prospect Theory parameters and the  optimal solution.

\subsection{Comparison with EUT}\label{Sec_CMP_EUT}
In this section, we focus on the traditional EUT case \cite{EUT}, which has been widely used in most prior literature (e.g., \cite{privacy_endogenous,approximately_auction_cost_data,buyingwithoutverification,value_privacy,game_quality,selling_privacy_auction,inception,bidguard,reap,bilateral,nonmonetarygame,linearregressiongame,meanfieldERM,socialcorrelation}) when dealing with uncertainty. Basically, EUT explains an individual's decision by evaluating the expected outcome (or utility) under uncertainty without considering risk and loss attitudes.  EUT can be considered as a special case of Prospect Theory when we choose $\lambda=1$ and $\beta=1$ in (\ref{valuation}). We compare the result of the data collector's approximated optimal solution under the EUT case with that under the general Prospect Theory case (excluding the special case of EUT).

\begin{Cor}\label{Cor_EUT_cmp}
	The data collector's approximated optimal $\tilde{\epsilon}^*_e$ under the EUT case is higher than that under the general Prospect Theory case.
\end{Cor}

We provide the proof of Corollary \ref{Cor_EUT_cmp} in  Appendix B. We conclude that compared with traditional EUT modeling, the data collector should adopt a more conservative privacy-preserving mechanism when considering the individuals' loss attitude predicted by the Prospect Theory. Without doing so, she will suffer a utility loss for not properly capturing individuals' behavioral characteristics.

\subsection{Impact of Prospect Theory Parameters}\label{sec_PTparameter}

To better understand the insights from Theorem \ref{Theorem_optimalsolution}, we study the impact of Prospect Theory parameters on the optimal differentially private mechanism in this section. Note that these parameters are intrinsic properties of individuals that the data collector can not control. The data collector needs to design the mechanism based on individuals' prospect properties.

\subsubsection{Impact of the Loss Aversion Parameter $\lambda$}
We first look at the impact of the Prospect Theory parameters $\lambda$  under the case of a zero reference point.

\begin{Cor}\label{Cor_lambda}
	The data collector's optimal $\tilde{\epsilon}^*$ decreases in $\lambda$.
\end{Cor}

We provide the proof of Corollary \ref{Cor_lambda} in Appendix C. Corollary \ref{Cor_lambda} suggests that  if the individuals are more loss averse (i.e., with a larger loss aversion parameter $\lambda$), they are less likely to participate in the data collection due to serious concerns of privacy loss. In order to attract individuals' participation, the data collector needs to adopt a more conservative privacy-preserving mechanism  to alleviate the concerns.

\subsubsection{Impact of the Reference Point $\epsilon_{ref}$}
Recall that a positive reference point means that individuals are tolerant.  They would perceive a privacy level outcome  as a gain if it is better than the reference even though it is not a perfect protection. Intuitively, the data collector can take advantage of individuals' tolerance towards privacy, and adopt a less conservative privacy-preserving mechanism comparing with the case of a zero reference point. However, is this intuition always true? Next, we will try to answer this question.

We compare the case of a positive reference point with a zero reference point. The methodology of computing  the data collector's approximated optimal privacy-preserving mechanism  under a large population approximation is similar to that under the case of a zero reference point. The main difference lies in the characterization of the prospect privacy levels of both participation and non-participation.

Under the case of a positive reference point, the prospect privacy level of participation based on (\ref{prospect_level2}) is (with the subscript $pos$ and superscript $p$):
\begin{equation}\label{equ_participation_prospectlevel_positive}
\begin{aligned}
&\epsilon^{p}_{ pos}=\frac{t}{m}\sum_{i=1}^{t}\left(\epsilon_{ref}-\frac{i}{m}\epsilon\right)^{\beta}-\left(1-\frac{t}{m}\right)\lambda\sum_{i=t+1}^{m}\left(\frac{i}{m}\epsilon-\epsilon_{ref}\right)^{\beta}, \\& \text{for} \ \ t \ \ \text{satisfying} \ \ \frac{t}{m}\epsilon<\epsilon_{ref} \ \  \text{and} \ \ \frac{t+1}{m}\epsilon>\epsilon_{ref}.
\end{aligned}
\end{equation}
The first summation term of (\ref{equ_participation_prospectlevel_positive}) corresponds to the gain part, and the second summation term corresponds to the loss part.

The prospect privacy level of non-participation is (with the subscript $pos$ and superscript $n$):
\begin{equation}
\epsilon^{n}_{pos} = \epsilon^{\beta}_{ref}.
\end{equation}
 Individuals would always enjoy certain gain from non-participation due to a better privacy protection than the  positive reference level.

Let $U^{\prime a}_{pos}(\epsilon)$ (formulated below) be the approximated derivative of the objective function and  $f_{pos}(\epsilon)$  be  the polynomial numerator in $U^{\prime a}_{pos}(\epsilon)$. Comparing the polynomial part $f_{pos}(\epsilon)$ with $f(\epsilon)$ in (\ref{zero_poly}) under the case of a zero reference point, we can see that the term $g(\epsilon_{p})$ in (\ref{zero_poly}) is replaced by the term $g(\epsilon^{p}_{pos})-g(\epsilon^{n}_{pos})$ in (\ref{nonzero_poly}). Then the approximated optimal solution $\tilde{\epsilon}^{*}_{pos}$ under the case of a positive reference point is  the root of $f_{pos}(\epsilon)=0$ in the feasible set  $\{\epsilon:g(\epsilon^{n}_{pos})-g(\epsilon^{p}_{pos}) <W_{\max}\}$.

\begin{equation}\label{approx_derivative1_pos}
U^{\prime a}_{pos}(\epsilon)=\frac{4}{\left(\frac{W_{\max}+g(\epsilon^{p}_{pos})-g(\epsilon^{n}_{pos}))}{W_{\max}}\right)^3 \epsilon ^3 N^2}f_{pos}(\epsilon),
\end{equation}
and

\begin{multline}\label{nonzero_poly}
f_{pos}(\epsilon) =  \left(W_{\max}+\frac{g(\epsilon^{p}_{pos})-g(\epsilon^{n}_{pos})}{W_{\max}}\right)\times\\ \left(1+ \frac{kN}{4l}\epsilon^3\frac{g'(\epsilon^{p}_{pos})-g'(\epsilon^{n}_{pos})}{W_{\max}}\right)- \frac{g'(\epsilon^{n}_{pos})-g'(\epsilon^{p}_{pos})}{W_{\max}}\epsilon.
\end{multline}

We compare the data collector's optimal privacy-preserving mechanism under both the cases of a positive reference point and a zero reference point. Remind that $\tilde{\epsilon}^*$ is  the approximated optimal solution under the case of a zero reference point, i.e, $f(\tilde{\epsilon}^*)=0$.

\begin{The}\label{The_ref}
	Comparing $\tilde{\epsilon}^{*}_{pos}$ with $\tilde{\epsilon}^*$, we have the following result:
	\begin{itemize}
		\item When $f_{pos}(\tilde{\epsilon}^*)<0$ is true, $\tilde{\epsilon}^{*}_{pos}<\tilde{\epsilon}^*$.
		\item When $f_{pos}(\tilde{\epsilon}^*)>0$ is true, $\tilde{\epsilon}^*_{pos}>\tilde{\epsilon}^*$. 
		\item When $f_{pos}(\tilde{\epsilon}^*)=0$ is true, $\tilde{\epsilon}^*_{pos}=\tilde{\epsilon}^*$. 
	\end{itemize} 
\end{The} 

\begin{figure*}[t] 
	\begin{minipage}{0.48\linewidth}
		\centering\includegraphics[width=3in]{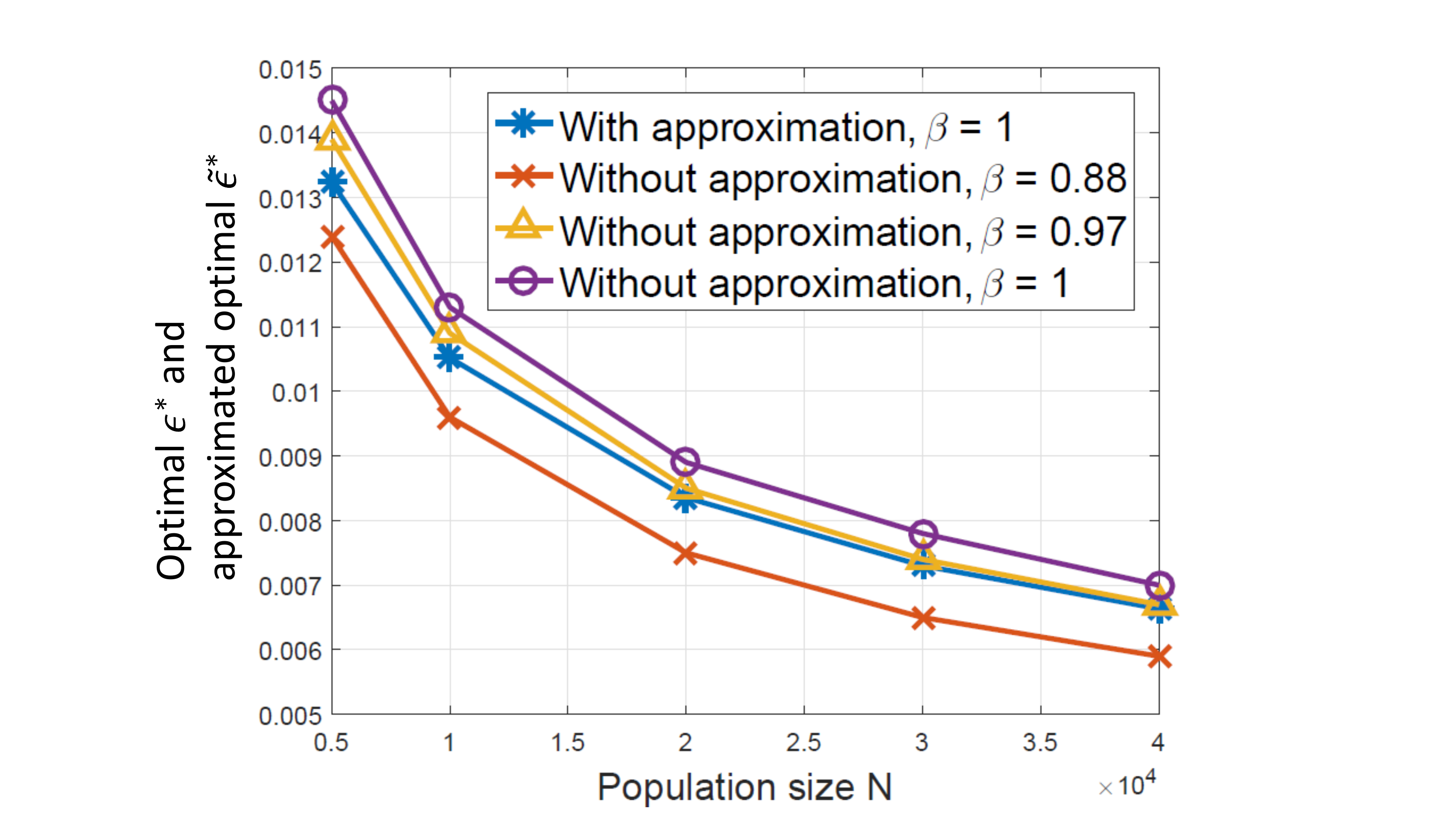}
		\caption{Comparison between optimal $\epsilon^*$ with and without approximation vs. $N$.}
		\label{N_appro}
	\end{minipage}
	\hfil
	\begin{minipage}{0.48\linewidth}
		\centering\includegraphics[width=2.3in]{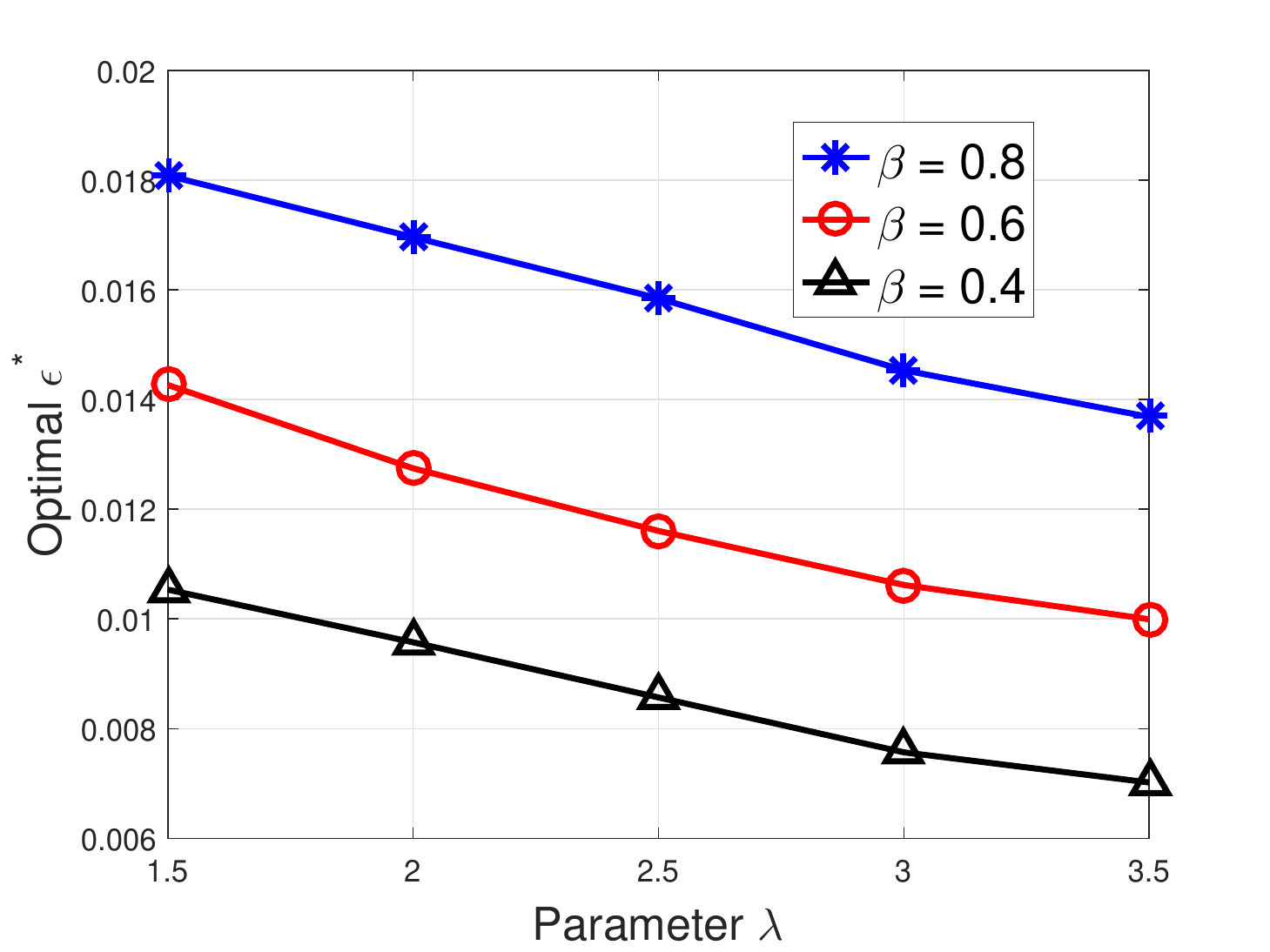}
		\caption{Optimal $\epsilon^*$ under different parameters $\lambda$ and $\beta$  for the zero reference point case.}
		\label{pro_PT}
	\end{minipage}
	\begin{minipage}{0.48\linewidth}
		\centering\includegraphics[width=2.3in]{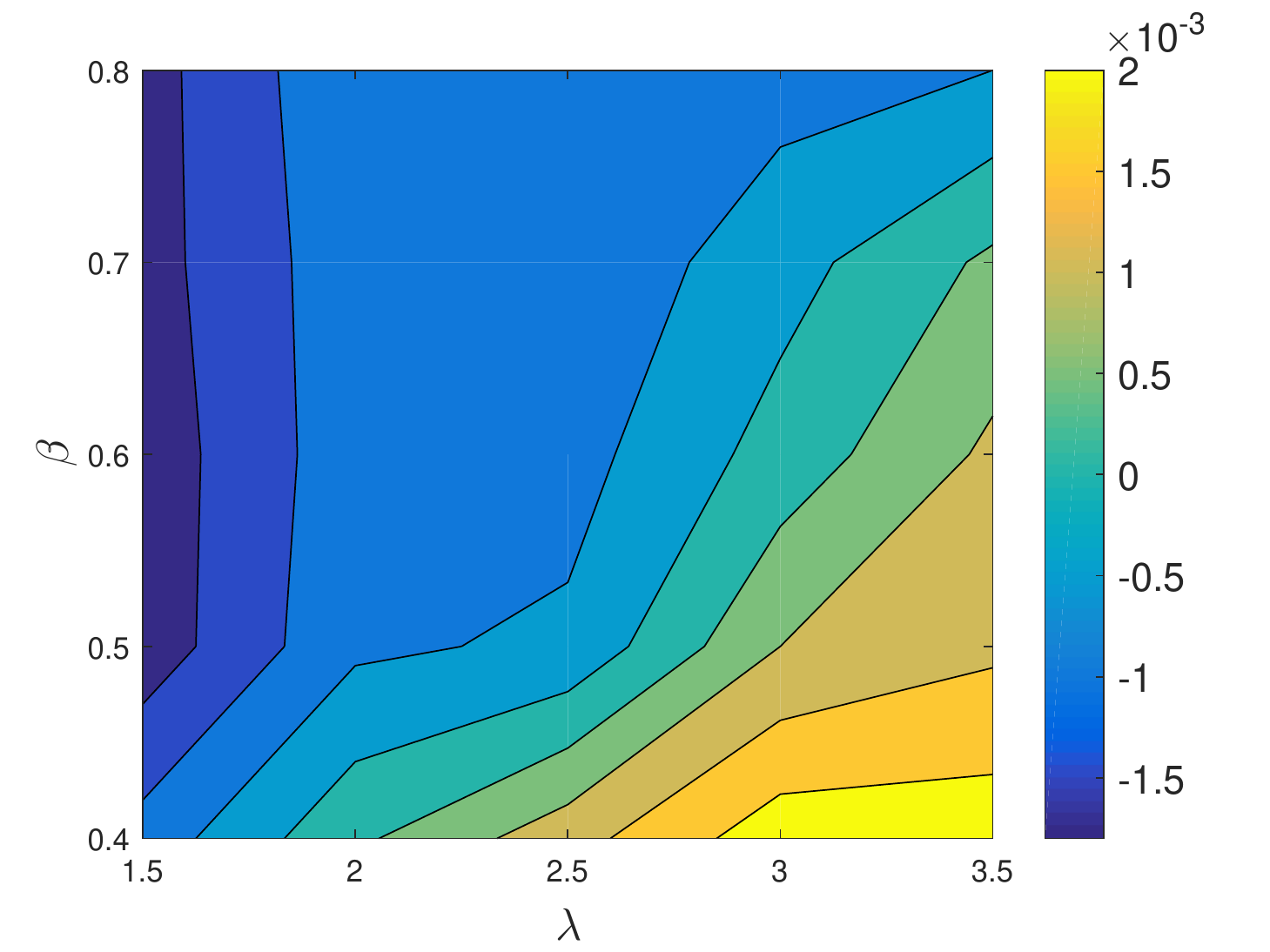}
		\caption{$\epsilon^*_{pos}-\epsilon^*$ under different parameters $\lambda$ and $\beta$.}
		\label{ref}
	\end{minipage}
	\hfil
	\begin{minipage}{0.48\linewidth}
		\centering\includegraphics[width=2.3in]{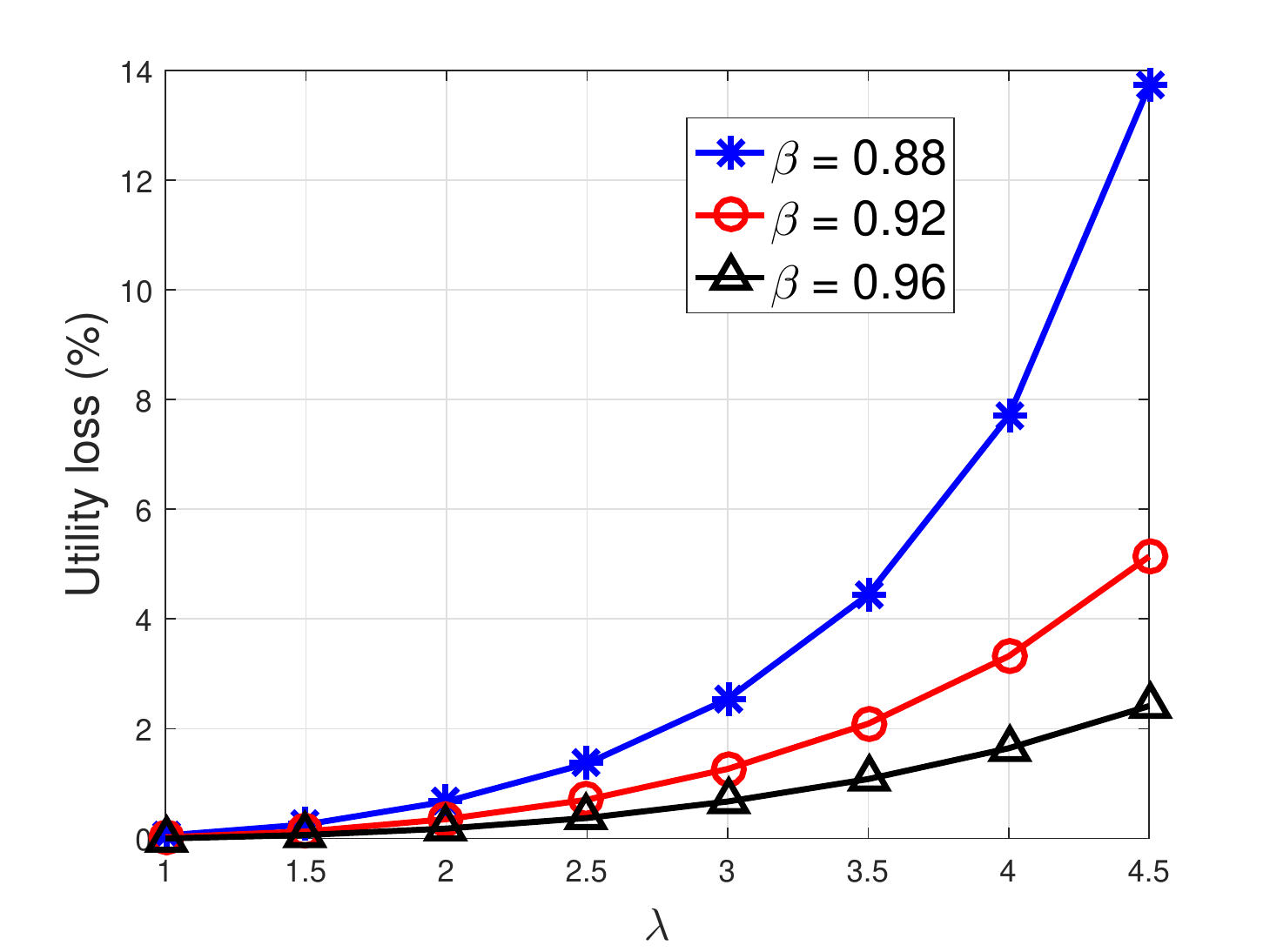}
		\caption{Utility loss (\%) under different $\lambda$ and $\beta$ for the zero reference point case.}
		\label{utility_cmp}
	\end{minipage}
\end{figure*}

We provide the proof of Theorem \ref{The_ref} in Appendix D. Theorem \ref{The_ref} indicates that a positive reference point (representing individuals' tolerance to the privacy issue) does not necessarily lead to a less  conservative privacy-preserving mechanism.  In contrast, the data collector needs to adopt a more conservative mechanism under a certain condition, i.e., $f_{pos}(\tilde{\epsilon}^*)<0$, under the case of a positive reference point.  More specifically, the condition $f_{pos}(\tilde{\epsilon}^*)<0$ holds if the loss aversion parameter $\lambda$ is relatively small or the risk parameter $\beta$ is relatively high. Both indicate that individuals are not sensitive to loss. When the reference point $\epsilon_{ref}$ changes from zero to positive, the perceived privacy protection gain from non-participation is more significant and overweights the loss reduction of participation. So individuals would prefer  not to participate. In this case, the data collector needs to enforce a more conservative privacy protection to encourage the individuals to participate.

\section{Numerical results and insights}\label{Simulation}

In this section, we evaluate the proposed privacy-preserving data collection mechanism from a variety of perspectives, including the  approximation performance, the impact of accurate prospect theoretic modeling, and the impact of Prospect Theory parameters.

\subsection{The Accuracy of the Approximated Solution}
First, we compare the optimal solution with and without approximation under different values of population size $N$ and parameter $\beta$. The result with approximation is calculated according to Theorem \ref{Theorem_optimalsolution}, and the result without approximation is obtained by an exhaustive search.  We change the population size $N$, and we compare the approximated solution $\tilde{\epsilon}^*$ under $\beta=1$ with the optimal solution $\epsilon^*$ under different values of $\beta$.

Fig. \ref{N_appro} shows that both the approximated and the optimal solutions decrease in $N$. This is because a larger number of participants can potentially provide a higher accuracy in data computation, which reduces the data collector's accuracy penalty due to the added noise. To attract more participants, the data collector would prefer to adopt  a more conservative privacy-preserving mechanism. \footnote{On the other hand, this does not mean that the data collector would include \emph{all} the individuals. Since the optimal $\epsilon^*$ is non-zero, i.e., the privacy protection is  imperfect, there may always exist some individuals who are reluctant to participate. The data collector needs to trade-off the balance between data amount benefit and accuracy penalty by optimizing the value of $\epsilon^*$.}

Comparing the two curves under  $\beta=1$, we see that the gap due to the approximation is relatively small, and such a gap decreases in $N$. For example, as $N$ changes from $4000$ to $40000$, the relative difference between $\tilde{\epsilon}^*$ with approximation and $\epsilon^*$ without approximation decreases from $8.7\%$ to $5.2\%$. This is because when $N$ becomes larger, the large population approximation becomes more accurate.

\subsection{The Impact of Prospect Theory Parameters}

In this subsection, we study the impact of Prospect Theory parameters $\lambda$, $\beta$, and the reference point $\epsilon_{ref}$. We consider a general truncated normal distribution \cite{truncated_normal} for the individuals' reward valuation. The uniform distribution  used in the theoretic analysis is a special case of truncated normal distribution when the variance approaches infinity. In this and the later simulations, we use the optimal solution $\epsilon^*$ \emph{without approximation} as the measurement of privacy-preserving mechanism. We will see that the simulation results match the theoretical results that we obtained \emph{through approximation} in Section \ref{Solving}.

We first focus on parameters $\lambda$ and $\beta$ for a zero reference point case. Fig. \ref{pro_PT} shows that the data collector's optimal solution $\epsilon^*$ decreases in the loss aversion parameter $\lambda$ and increases in the risk aversion parameter $\beta$. Intuitively,   the loss aversion parameter $\lambda$ is larger or the parameter $\beta$ is smaller, individuals would subjectively experience more serious privacy loss  once choosing participation. So the data collector needs to adopt a more conservative privacy-preserving mechanism to encourage individuals' participation.

We then focus on the reference point $\epsilon_{ref}$. A higher $\epsilon_{ref}$ means that individuals are more tolerant about their privacy loss. We set the positive reference point $\epsilon_{ref}$ to $0.01$. Fig.\ref{ref} shows the difference of optimal solution between the case of a positive reference point and a zero reference point (i.e., $\epsilon^*_{pos}-\epsilon^*$) under different values of parameters $\lambda$ and $\beta$. 

We first focus on the bottom right region in Fig. \ref{ref} where $\lambda$ is large (i.e., $\lambda \geq 2$) and $\beta$ is small( i.e., $\beta \leq 0.5$). The difference $\epsilon^*_{pos}-\epsilon^*$ is positive in this region, i.e., the data collector offers a less conservative privacy-preserving mechanism with a positive reference point. This is because in this region individuals are very sensitive to loss  (see Fig. \ref{valuationfunction_1}). When $\epsilon_{ref}$ increases from zero, a participant is less likely to experience loss, hence the subjectively perceived privacy cost from participation significantly decreases. Such a  privacy loss tolerance attitude allows the data collector to adopt a less conservative privacy-preserving mechanism.

Next, we consider the upper left region in Fig. \ref{ref} where $\lambda$ is small(i.e., $\lambda \leq 1.5$)  or $\beta$ is large (i.e., $\beta > 0.5$). The difference $\epsilon^*_{pos}-\epsilon^*$ is negative in this region, i.e., the data collector offers a more conservative privacy-preserving mechanism with a positive reference point. This is because in this region individuals are less sensitive to loss  (see Fig. \ref{pro_valuation}). So when $\epsilon_{ref}$ increases from zero,  the reduction of privacy protection loss from participation is less significant compared with privacy protection gain from non-participation. Hence the data collector needs to enforce a more conservative privacy protection to encourage the individuals to participate.

\subsection{The Impact of Prospect Theoretic Modeling Accuracy}

In this subsection, we show the importance when using an accurate prospect theoretic modeling. In Section \ref{Sec_CMP_EUT}, Corollary \ref{Cor_EUT_cmp} shows that a data collector should adopt a more conservative privacy-preserving mechanism when considering the prospect theoretic characteristics of individuals. However, if the data collector assumes that individuals make decisions based on EUT (while the actual decisions are made based on Prospect Theory),  she can suffer a significant utility loss.

Fig. \ref{utility_cmp} shows the relative  utility loss (normalized by the maximum utility) when the mismatch happens  under different Prospect Theory parameters $\lambda$ and $\beta$ for a zero reference point case.  We see that the loss increases significantly  when the parameters $\lambda$ and $\beta$ deviate from EUT case (which corresponds to  $\lambda=1$ and $\beta=1$). For example,  when $\lambda=4.5$ and $\beta=0.88$, the utility loss is about  14\%. This indicates the importance for the data collector to have an accurate estimation of the users' Prospect Theory parameters through extensive data collection and analysis.

\section{Impact of the Heterogeneity of Individuals}\label{sec_heterogeneity}

Previous analysis and simulations in Sections \ref{Solving} and  \ref{Simulation} are based on the assumption of homogeneous individuals.  Here we numerically study  a more realistic situation where different individuals may have different behavior characterizations \cite{experienced,tractable,lossaversion,gender}, and explore  the impact of such heterogeneity on the data collector's optimal privacy-preserving mechanism. We will leave the analytical study of the heterogeneous parameter model in the future work.

\begin{figure}[!t]	
	\centering
	\subfloat[$\lambda\sim$ Gamma (3.24,0.60).]{\includegraphics[width=0.52\linewidth]{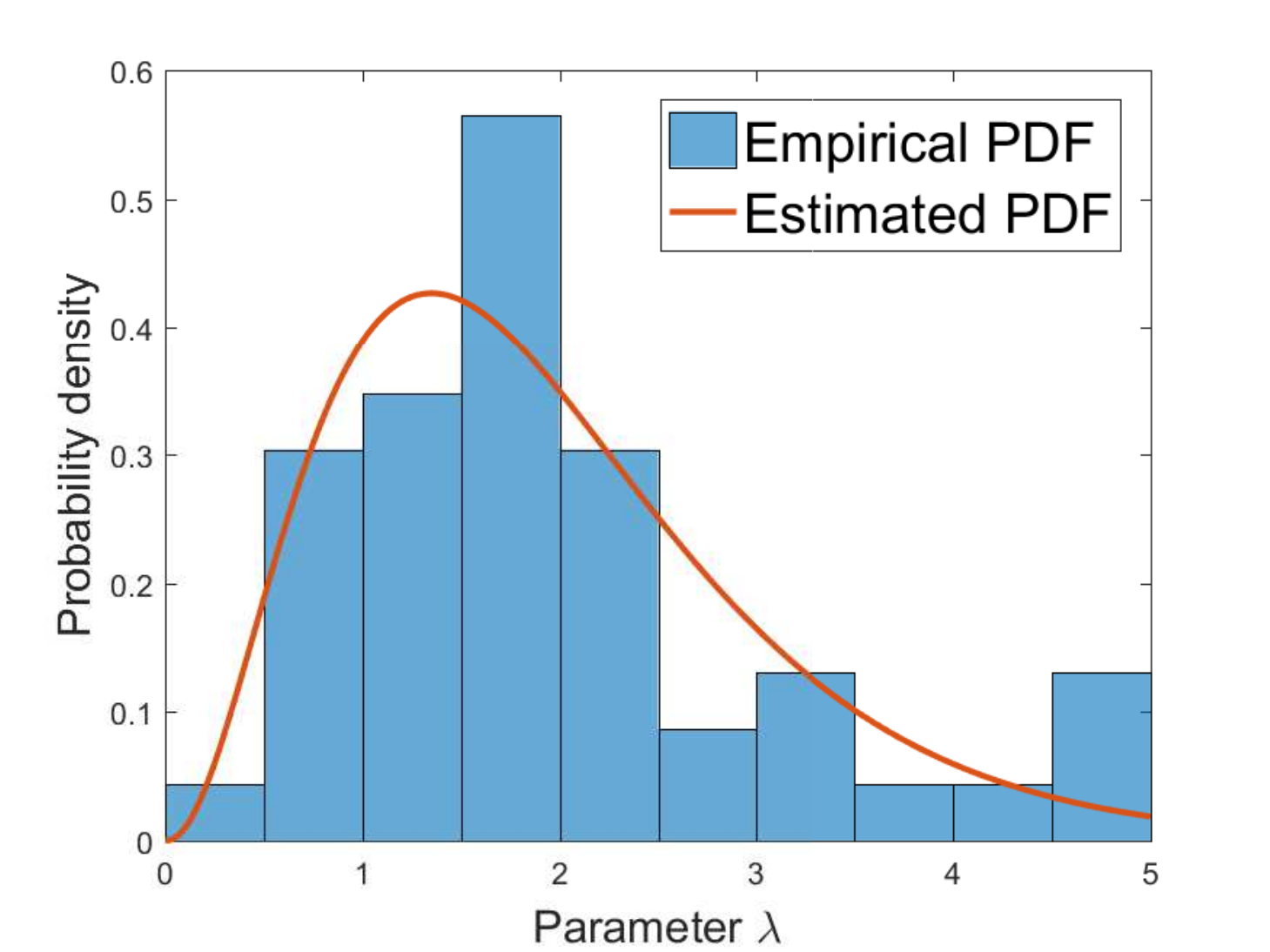}\label{Gamma_lambda}}
	\hfil
	\subfloat[$\beta \sim$ Gamma (12.87,0.06).]{\includegraphics[width=0.52\linewidth]{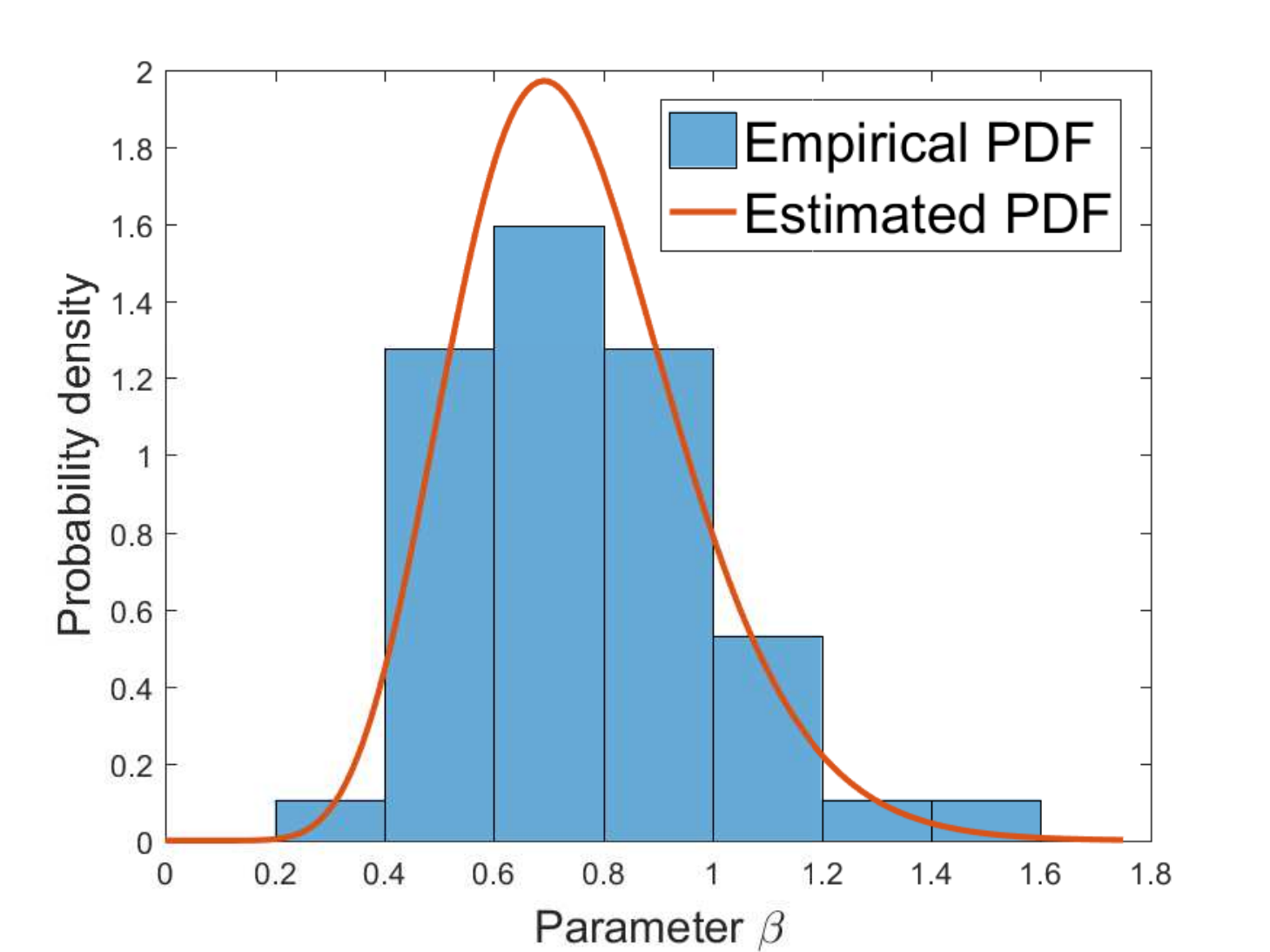}\label{Gamma_beta}}
	\caption{Probability density of parameters $\lambda$ and $\beta$ based on the data from \cite{tractable}.}
	\label{PT_hist}
	\vspace{-3mm}
\end{figure}

We first want to understand the distribution of Prospect Theory  parameters in real life.  The data comes from  the literature in the area of psychology and behavioral economics that investigated the Prospect Theory parameters of each subject in the experiments (e.g., \cite{experienced,tractable,lossaversion,gender}). More specifically, we utilize the experimental results in \cite{tractable}.\footnote{The experiments are based on subjects' reactions under monetary reward, instead of privacy protection. Monetary reward is widely used in  psychological experiments  and the corresponding literature. The purpose of utilizing these reported data in the literature is to provide a relative realistic context in terms of the Prospect Theory parameter choices, instead of randomly generating these parameters. Our theoretical results apply for any Prospect Theory parameter settings, and it is important future work to perform field studies to understand the actual parameters in the privacy preserving contexts for populations of different age, sex, education background, and countries.  } Fig. \ref{PT_hist} shows  the empirical probability density of parameters $\beta$ and $\lambda $ based on the data from  \cite{tractable}. We perform data fitting by using the chi-squared test \cite{simulation}, we conclude that the parameter $\lambda$  follows a Gamma distribution with a shape parameter $k_{\lambda}=3.2433$ and a scale parameter  $\theta_{\lambda}=0.6018$ ($p=0.4768$), and that the parameter $\beta$ follows a Gamma distribution with a shape parameter $k_{\beta}=12.8662$ and a scale parameter $\theta_{\beta}=0.0583$ ($p = 0.1278$).\footnote{The  mean of the Gamma distributed random variable is the product of the shape parameter $k$ and the scale parameter $\theta$, i.e., $k\cdot\theta$. }

Utilizing the above empirical data, we will study the impact of the heterogeneity of parameter $\lambda$. We generate this parameter through the Gamma distribution with a fixed mean ($\mu_{\lambda}=k_{\lambda}\times\theta_{\lambda}=1.95$) based on Fig. \ref{Gamma_lambda}  and different values of variance. We fix other parameters including the parameter $\beta=k_{\beta}\times\theta_{\beta}=0.75$ and the reference point $\epsilon_{ref}=0$.

\begin{figure}[t]
	\centering
	\includegraphics[width=2.3in]{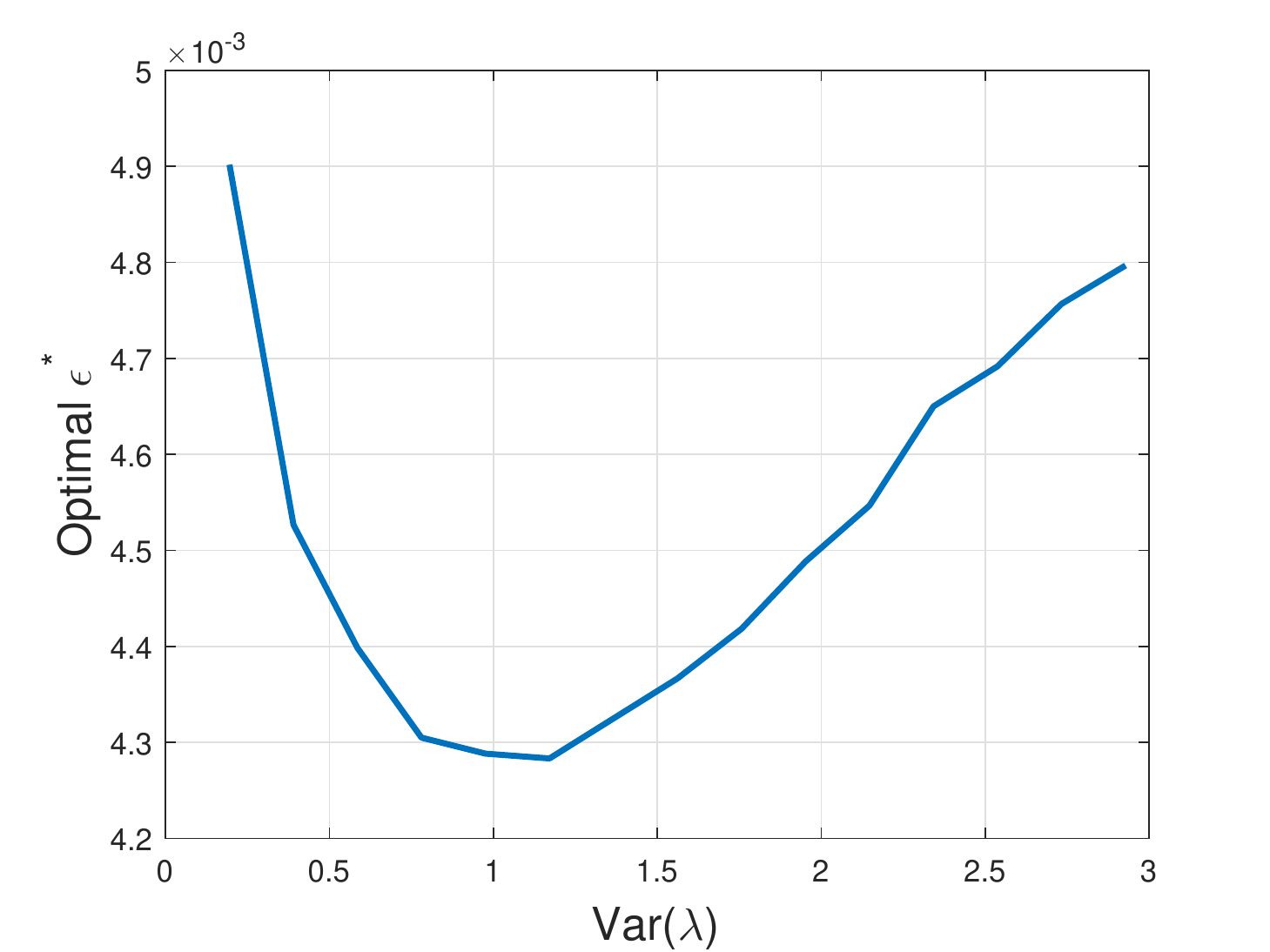}
	\caption{Impact of variance of $\lambda$ on optimal $\epsilon^*$.}
	\label{1}
\end{figure}

\begin{figure}[t]
	\centering
	\subfloat[$Var(\lambda)=0.1952,\epsilon^*=4.9\times 10^{-3}$.]{\includegraphics[width=0.6\linewidth]{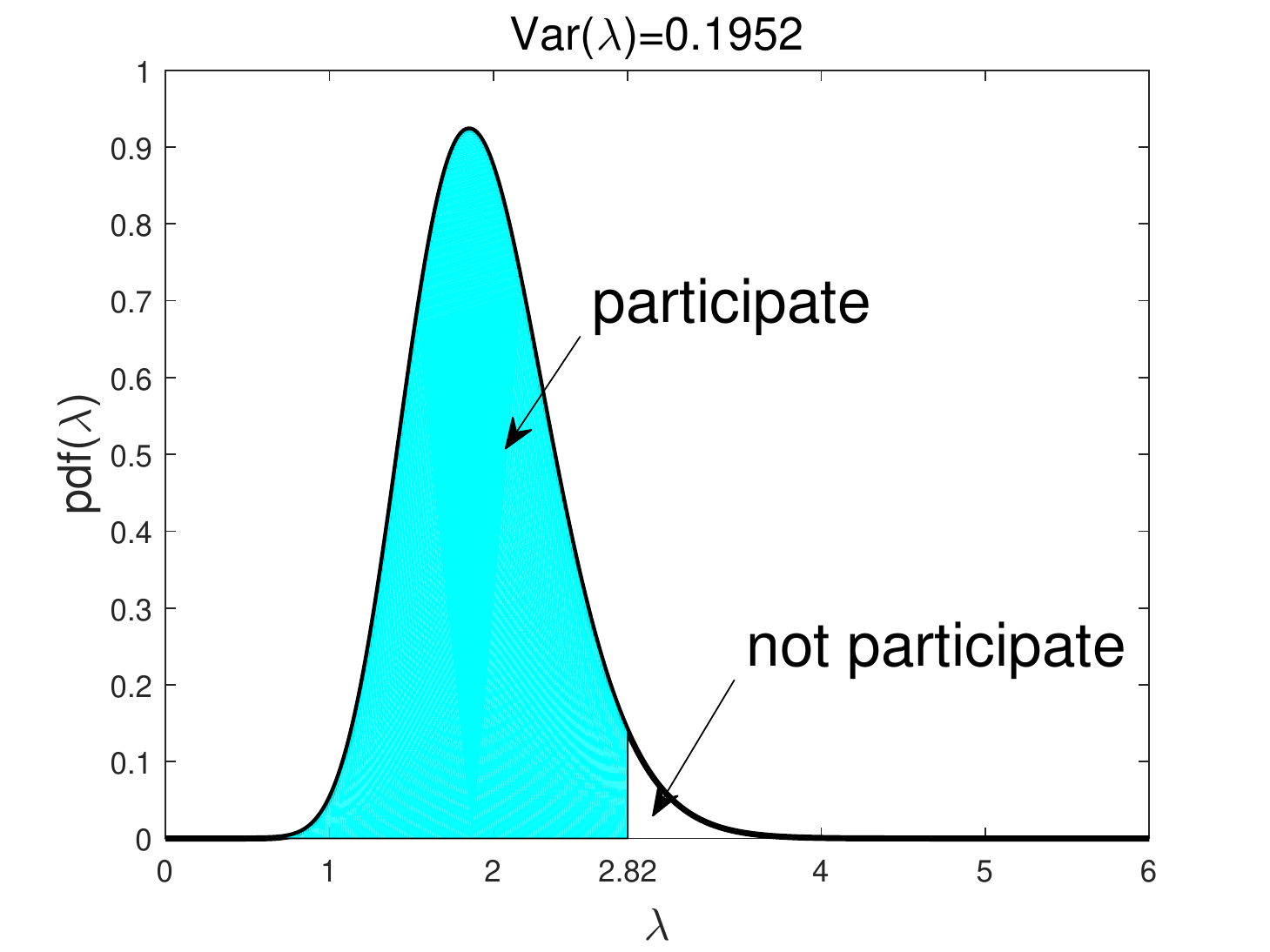}\label{11}}
	\hfil
	\subfloat[$Var(\lambda)=0.7807,\epsilon^*=4.3\times 10^{-3}$.]{\includegraphics[width=0.6\linewidth]{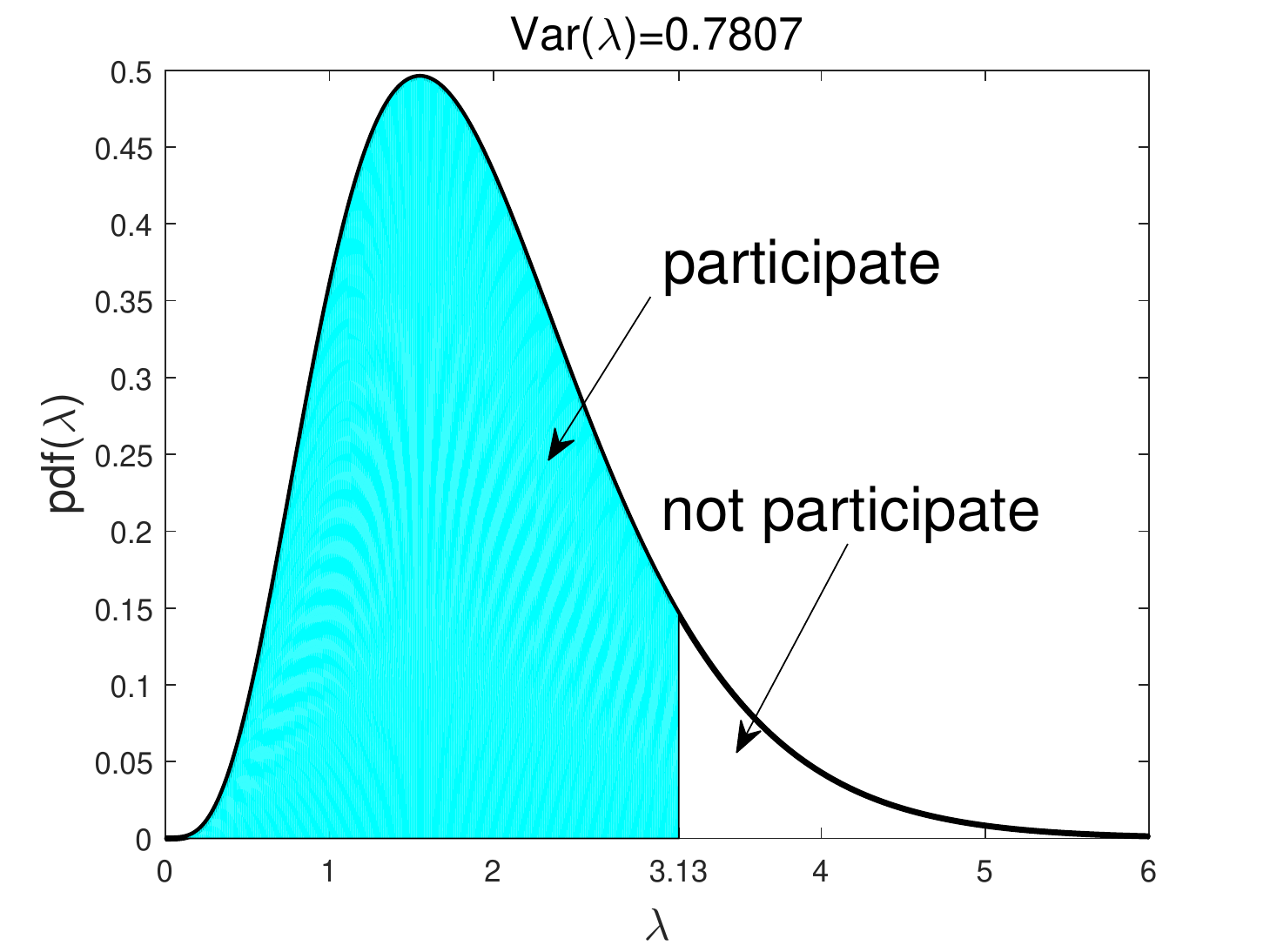}\label{22}}
	\hfil
	\subfloat[$Var(\lambda)=2.3421,\epsilon^*=4.7\times 10^{-3}$.]{\includegraphics[width=0.6\linewidth]{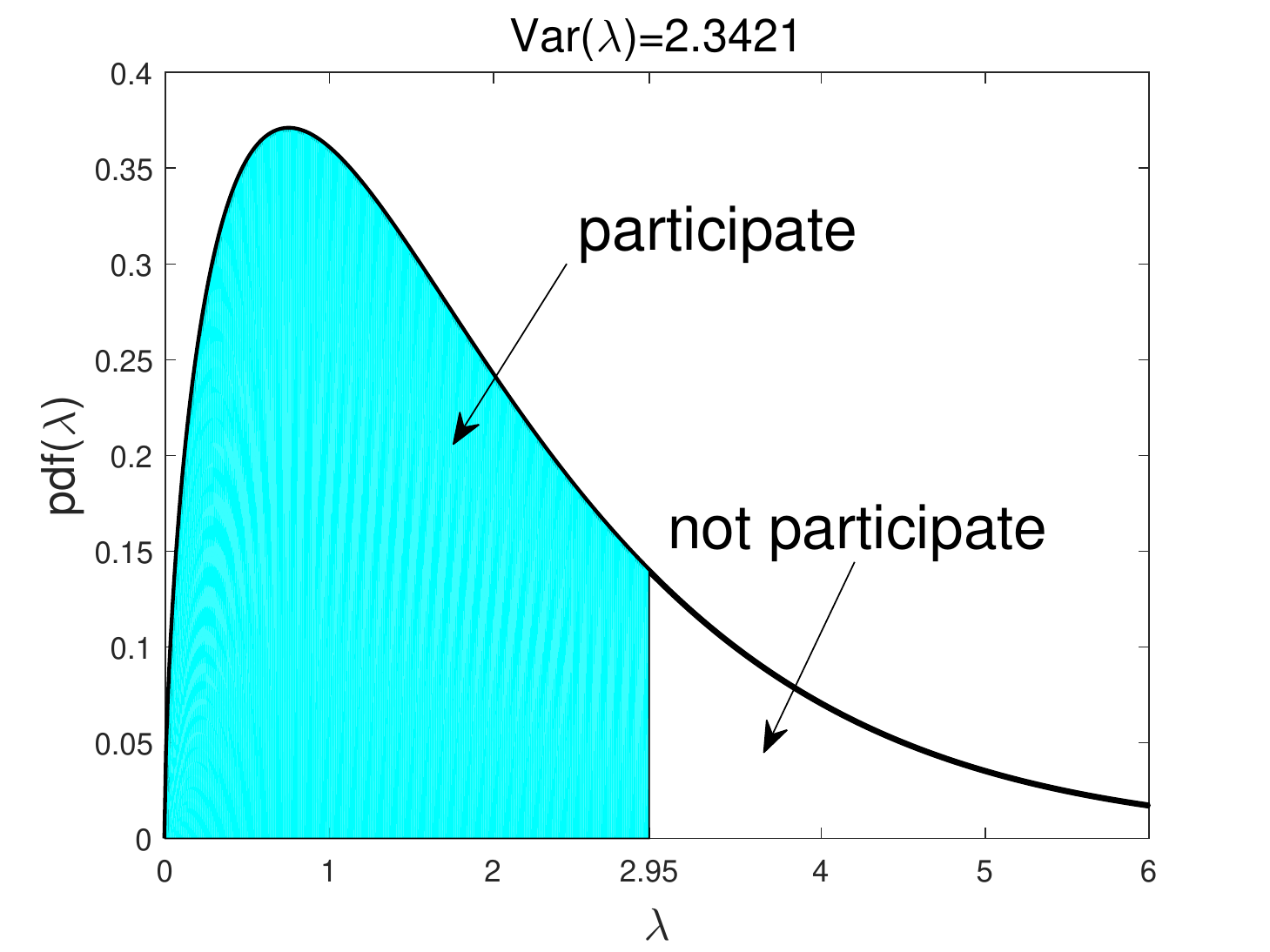}\label{33}}
	\caption{Participation under different variance of $\lambda$.}
	\label{participation_lambda}
\end{figure}

Fig. \ref{1} shows  how the data collector's optimal $\epsilon^*$ changes with the variance of $\lambda$. We can see that the optimal $\epsilon^*$ firstly decreases in the variance and then increases in the variance. To better understand such an impact, we visualize individuals' participation in Fig. \ref{participation_lambda} given the optimal $\epsilon^*$ under different variances of $\lambda$ (in an increasing order). Those with the lower value of $\lambda$ (less loss aversion) would participate and the corresponding threshold changes with the variance.

Combining Fig. \ref{1} and Fig.  \ref{participation_lambda}, we are able to illustrate the intuition. At a low diversity level (small variance  such as in Fig. \ref{11}), most individuals' $\lambda$ values are close to the mean. Comparing Fig. \ref{11} (small variance) and Fig. \ref{22} (medium variance), we can see that as the variance becomes larger, there are more individuals  with $\lambda$ values further  away from the mean. To attract those individuals with $\lambda$ values higher than the mean, the data collector needs to adopt a more conservative privacy-preserving mechanism  when the variance increases. A more conservative mechanism corresponds to a larger participation threshold (3.13) in Fig. \ref{22} than that (2.82) in Fig. \ref{11}. 

However, at a high diversity level  (as in Fig. \ref{33}), individuals' $\lambda$ values spread around a big range. Comparing Fig. \ref{22} (medium variance) and Fig. \ref{33} (large variance), we can see that as the variance becomes very large, more individuals have very high $\lambda$ values. Those individuals are more loss averse  and are difficult to be motivated to  participate. In this case, a more conservative mechanism to attract those individuals would result in more accuracy penalty due to a larger variance of the added noise. Instead, it would be better for the data collector to ignore those with very high $\lambda$ values and to consider a relatively less conservative mechanism.  A less conservative mechanism corresponds to a smaller participation threshold (2.95) in Fig. \ref{33} than that (3.13) in Fig. \ref{22}.

\begin{figure}[!t]
	\centering
	\includegraphics[width=2.3in]{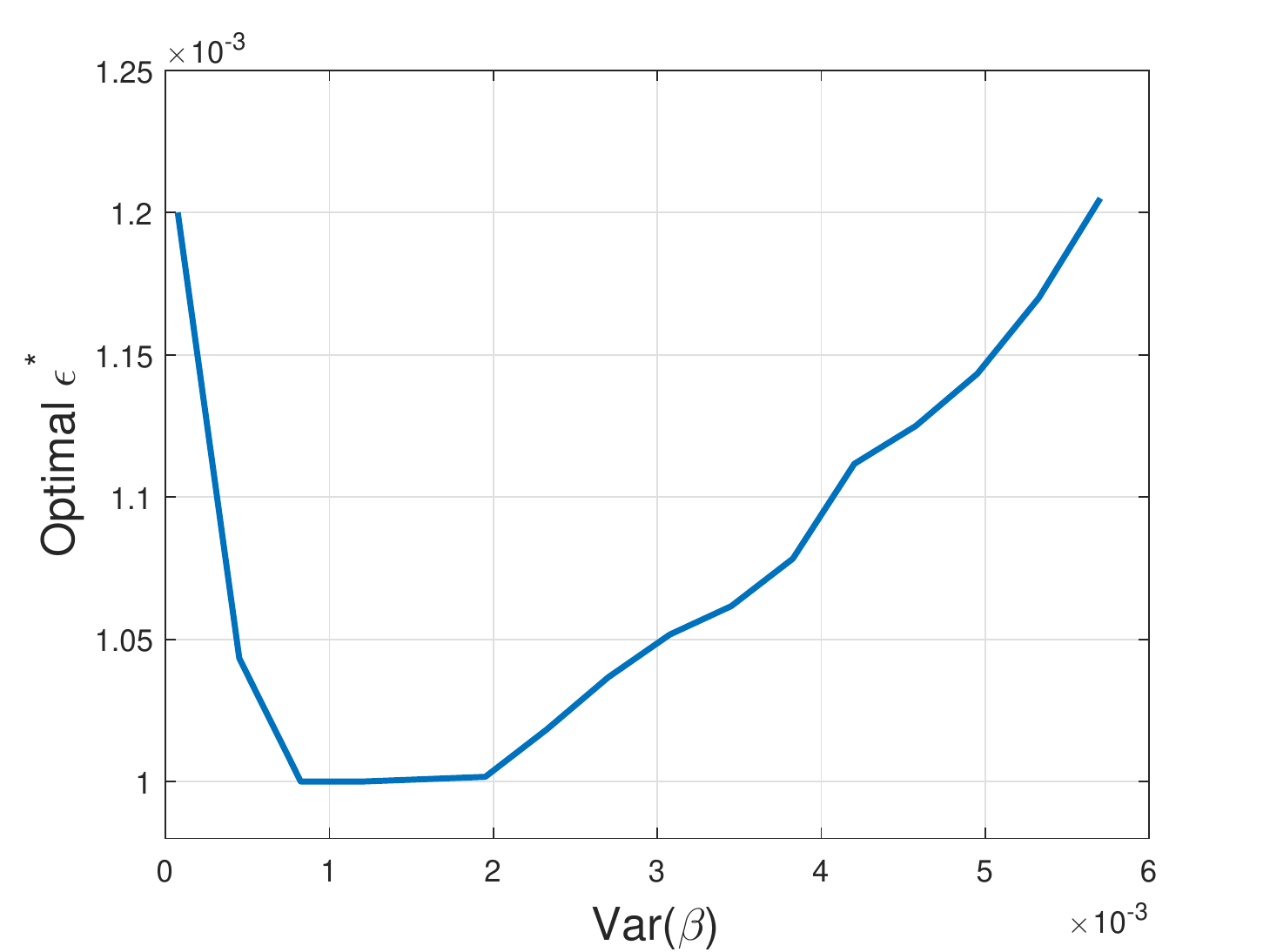}
	\caption{Impact of variance of $\beta$ on optimal $\epsilon^*$.}
	\label{3}
	\vspace{-5mm}
\end{figure}

We also study the impact of the heterogeneity of parameter $\beta$. Similarly, we generate the parameter $\beta$ through the Gamma distribution with a fixed mean ($\mu_{\beta}=k_{\beta}\times\theta_{\beta}=0.75$) based on Fig. \ref{Gamma_beta}  and different values of variance. We fix $\lambda=k_{\lambda}\times\theta_{\lambda}=1.95$ and $\epsilon_{ref}=0$. Fig. \ref{3} shows  the optimal $\epsilon^*$  under different variances of $\beta$.  The pattern is similar to that in Fig. \ref{1}: the optimal $\epsilon^*$ firstly decreases  and then increases in the variance of $\beta$. The corresponding insights are also the same as that from the parameter $\lambda$.

\section{Conclusion}\label{Conclusion}

In this paper, we analyzed a privacy-preserving data collection problem with the privacy protection uncertainty. To the best our knowledge, this is the first theoretical study of Prospect Theory application in the area of privacy protection.  We demonstrate the importance of a realistic and accurate individual decision modeling to the privacy-preserving mechanism design.  Considering the loss and risk attitudes based on prospect theoretic modeling, the data collector should adopt a more conservative privacy-preserving mechanism  compared with the one derived based on the expected utility theory modeling. Moreover, a more tolerant attitude on privacy loss predicted by a  positive reference point does not always indicate a less conservative mechanism.

For the future work, we will consider the case where the participants can misreport their data. For example,  the participant would like to protect his privacy on his own by reporting a noisy version of data. Considering both the risk aversion and loss aversion of the participants, the data collector needs to design an incentive mechanism that effectively induces truthful reporting from the users.

\ifCLASSOPTIONcaptionsoff
  \newpage
\fi

\bibliographystyle{IEEEtran}

\bibliography{ref}

\vspace{-5.25cm}
\begin{IEEEbiography}[{\includegraphics[width=1in,height=1.25in,clip,keepaspectratio]{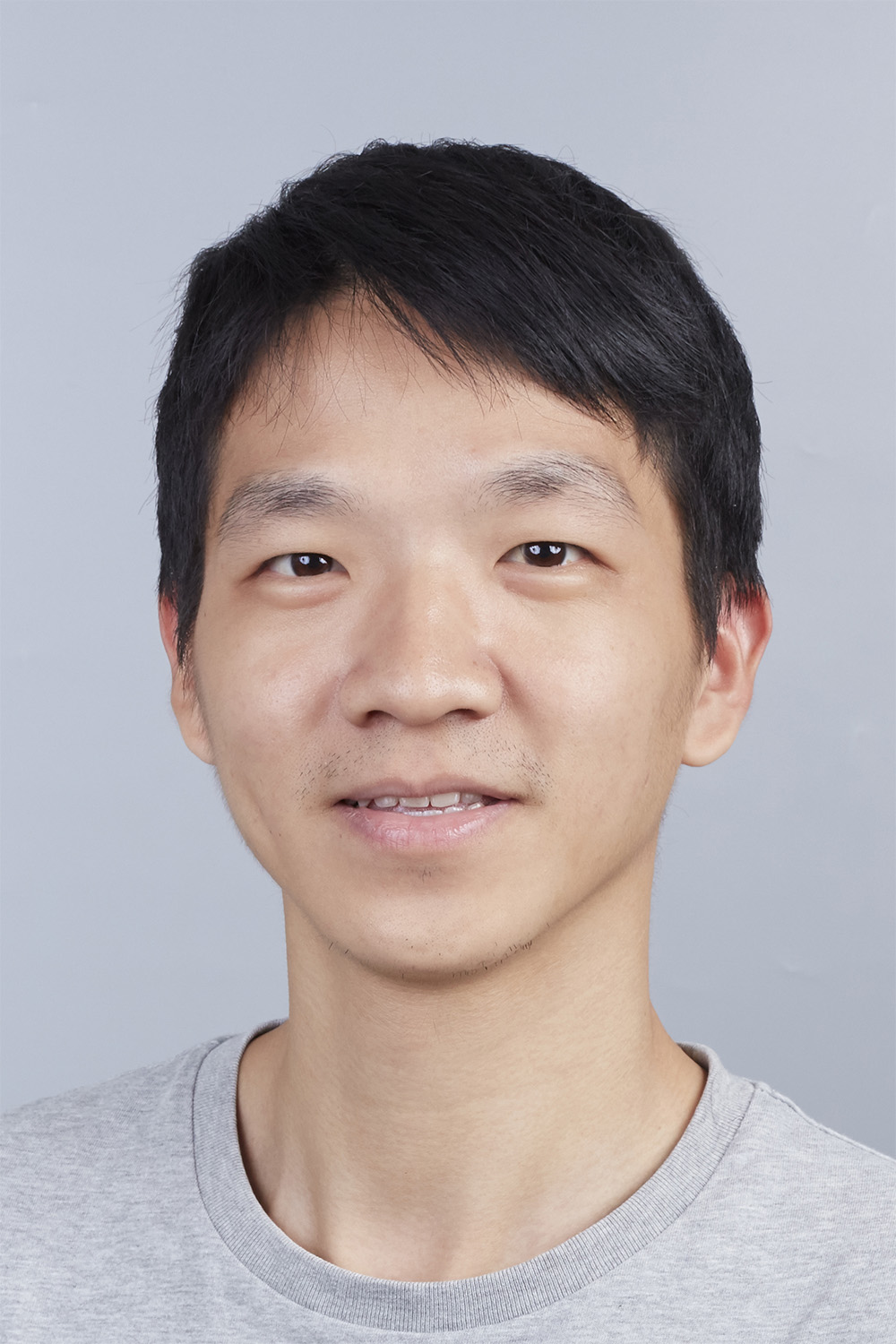}}]{Guocheng Liao} 
received the B.E. degree from Sun Yat-sen University in 2016. He is now pursuing the Ph.D. degree with Department of Information Engineering, The Chinese University of Hong Kong. His current research interests include data privacy and game theory. He received 2017 IEEE GLOBECOM ComSoc Young Professional Best Paper Award. 
\end{IEEEbiography}
\vspace{-5.1cm}
\begin{IEEEbiography}[{\includegraphics[width=1in,height=1.25in,clip,keepaspectratio]{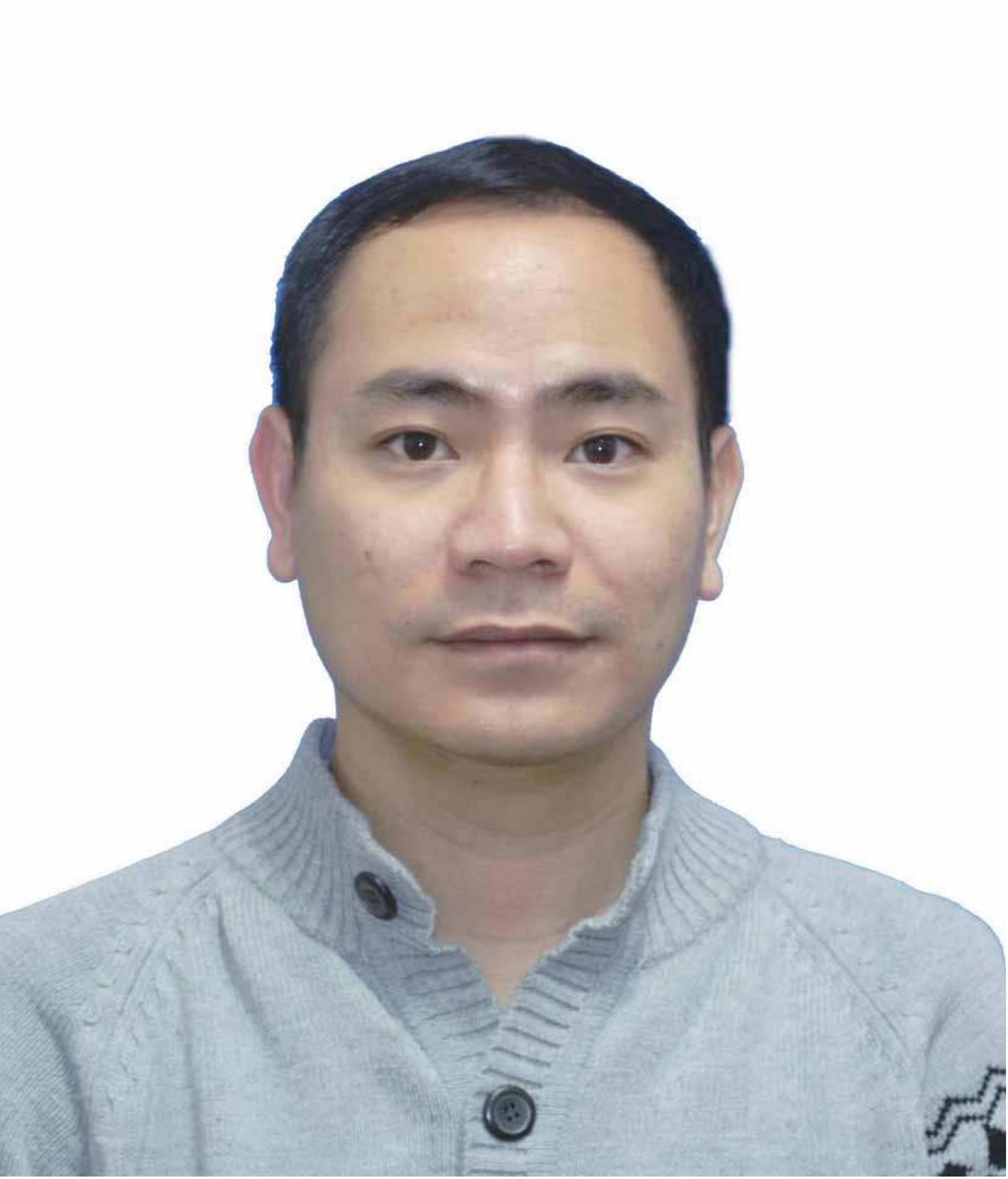}}]{Xu Chen} received the Ph.D. degree in information engineering from the Chinese University of Hong Kong in 2012. He is a Full Professor with Sun Yat-sen University, Guangzhou, China, and the Vice Director of the National and Local Joint Engineering Laboratory of Digital Home Interactive Applications. He was a Post-Doctoral Research Associate with Arizona State University, Tempe, USA, from 2012 to 2014, and a Humboldt Scholar Fellow with the Institute of Computer Science, University of Goettingen, Germany, from 2014 to 2016. He was a recipient of the Prestigious Humboldt Research Fellowship awarded by Alexander von Humboldt Foundation of Germany, the 2014 Hong Kong Young Scientist Runner-Up Award, the 2016 Thousand Talents Plan Award for Young Professionals of China, the 2017 IEEE Communication Society Asia–Pacific Outstanding Young Researcher Award, the 2017 IEEE ComSoc Young Professional Best Paper Award, the Honorable Mention Award of 2010 IEEE international conference on Intelligence and Security Informatics, the Best Paper Runner-Up Award of 2014 IEEE International Conference on Computer Communications (INFOCOM), and the Best Paper Award of 2017 IEEE International Conference on Communications. He is currently an Area Editor of IEEE Open Journal of the Communications Society, an Associate Editor of the IEEE Transactions Wireless Communications, IEEE Internet of Things Journal and IEEE Journal on Selected
	Areas in Communications (JSAC) Series on Network Softwarization and Enablers.
\end{IEEEbiography}
\vspace{9cm}

\begin{IEEEbiography}[{\includegraphics[width=1in,height=1.25in,clip,keepaspectratio]{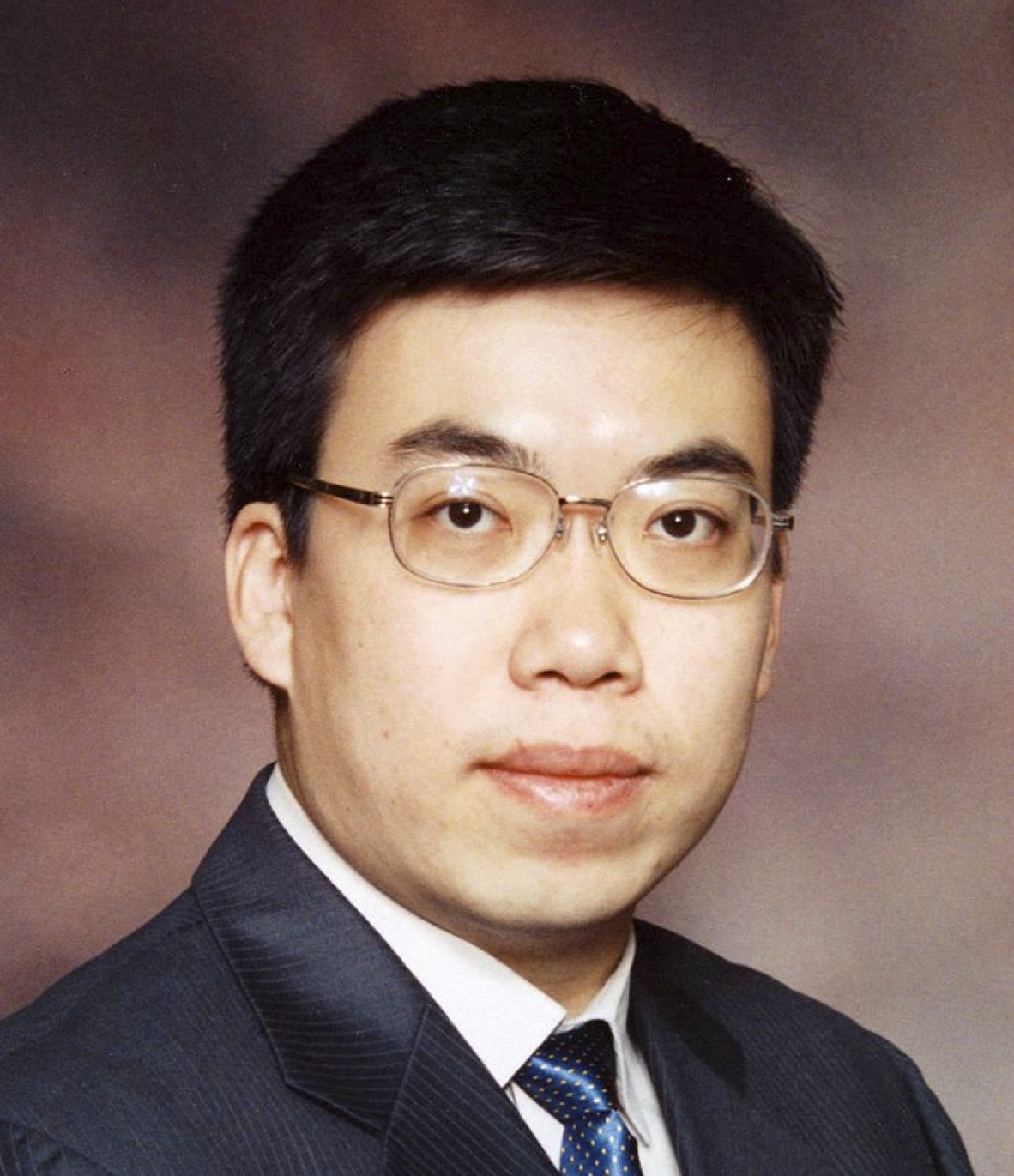}}]{Jianwei Huang}  is a Presidential Chair Professor and the Associate Dean of the School of Science and Engineering, The Chinese University of Hong Kong, Shenzhen. He is also the Associate Director of Shenzhen Institute of Artificial Intelligence and Robotics for Society, and a Professor in the Department of Information Engineering, The Chinese University of Hong Kong. He received the Ph.D. degree from Northwestern University in 2005, and worked as a Postdoc Research Associate at Princeton University during 2005-2007. He has been an IEEE Fellow, a Distinguished Lecturer of IEEE Communications Society, and a Clarivate Analytics Highly Cited Researcher in Computer Science. He is the co-author of 9 Best Paper Awards, including IEEE Marconi Prize Paper Award in Wireless Communications in 2011. He has co-authored six books, including the textbook on "Wireless Network Pricing." He received the CUHK Young Researcher Award in 2014 and IEEE ComSoc Asia-Pacific Outstanding Young Researcher Award in 2009. He has served as an Associate Editor of IEEE Transactions on Mobile Computing, IEEE/ACM Transactions on Networking, IEEE Transactions on Network Science and Engineering, IEEE Transactions on Wireless Communications, IEEE Journal on Selected Areas in Communications - Cognitive Radio Series, and IEEE Transactions on Cognitive Communications and Networking. He has served as an Editor of Wiley Information and Communication Technology Series, Springer Encyclopedia of Wireless Networks, and Springer Handbook of Cognitive Radio. He has served as the Chair of IEEE ComSoc Cognitive Network Technical Committee and Multimedia Communications Technical Committee. He is the Associate Editor-in-Chief of IEEE Open Journal of the Communications Society. He is the recipient of IEEE ComSoc Multimedia Communications Technical Committee Distinguished Service Award in 2015 and IEEE GLOBECOM Outstanding Service Award in 2010. More detailed information can be found at http://jianwei.ie.cuhk.edu.hk/.
\end{IEEEbiography}

\clearpage
\section*{Appendix}

\subsection{Proof of Theorem \ref{Theorem_optimalsolution}}\label{sec_A}
The proof consists of three steps. We first derive the number of participants in Stage \uppercase\expandafter{\romannumeral2} as a function of $\epsilon$. We then obtain the data collector's utility as a function of $\epsilon$ and its derivate. We finally approximate the derivative and obtain the unique root in the feasible set, which is the approximated optimal solution.
 
 \textbf{Step 1}:  Based on (\ref{participants_number}) and (\ref{reward_valuation_distribution}), we compute the number of participants among all individuals as a function of $\epsilon$:

\begin{equation}\label{participants_number_uniform}
n(\epsilon)=\begin{cases}
N\frac{W_{\max}+g(\epsilon_p)}{W_{\max}}, & \quad \text{if} \ -g(\epsilon_p)< W_{\max};\\
0, & \quad \text{otherwise}.
\end{cases} 
\end{equation}
Recall that $N$ is the number of all individuals. Here $g(\epsilon_p)=-M\epsilon^\beta$ based on (\ref{prospect_level2}) where $M = c\lambda/m(1/m)^{\beta}\sum_{i=1}^{m}i^\beta$. Since $\beta=1$ we have $g(\epsilon_p)=-M\epsilon$ where $M = c\lambda (m+1)/(2m)$.

\textbf{Step 2}: We derive the data collector's utility as a function of $\epsilon$ and its derivate. We consider the nontrivial case $-g(\epsilon_p)< W_{\max}$, such that there always exists some individuals who would like to participate. So we have the feasible set:  
\begin{equation}\label{equ_feasibleset}
\{\epsilon: -g(\epsilon_p) < W_{\max}\}.
\end{equation}
Recall that $\epsilon_{p}$ is the prospect privacy level of participation under the $\epsilon$-differentially private mechanism. The data collector's utility function and its derivative are as follows:

\begin{equation}
{U_c(\epsilon)}=1-\frac{k}{1+l N\frac{W_{\max}+g(\epsilon_p)}{W_{\max}}}-\frac{2}{N^2\epsilon^2\left(\frac{W_{\max}+g(\epsilon_p)}{W_{\max}}\right)^2,} 
\end{equation}
and

\begin{equation}\label{equ_derivative}
U'_c(\epsilon) = \frac{klN\frac{g'_p(\epsilon)}{W_{\max}}}{\left[1+lN\frac{W_{\max}+g(\epsilon_p)}{W_{\max}}\right]^2}+ \frac{4}{N^2}\frac{\frac{W_{\max}+g(\epsilon_p)+g'(\epsilon_p)\epsilon}{W_{\max}}}{\left[\frac{W_{\max}+g(\epsilon_p)}{W_{\max}}\right]^3\epsilon^3}. 
\end{equation}

\textbf{Step 3}: We approximate the derivative to find the approximated optimal solution. We approximate  $1+l N\left(W_{\max}+g(\epsilon_p)\right)/(W_{\max})$ with $l N\left(W_{\max}+g(\epsilon_p)\right)/(W_{\max})$, considering a large population size $N$. Then we can obtain an approximated (denoted by the superscript $a$) version of (\ref{equ_derivative}):
\begin{equation}\label{approx_derivative2}
U^{\prime a}_{c}(\epsilon)=\frac{4}{\left(\frac{W_{\max}+g(\epsilon_p)}{W_{\max}}\right)^3 \epsilon ^3 N^2}f(\epsilon).
\end{equation}
Here
\begin{equation}
f(\epsilon) = \left(\frac{W_{\max}+g(\epsilon_p)}{W_{\max}}\right)\left(1+ \frac{kN}{4l}\epsilon^{3}\frac{g'(\epsilon_p)}{W_{\max}}\right)+\epsilon\frac{g'(\epsilon_p)}{W_{\max}}, 
\end{equation}
where $g(\epsilon)=-M\epsilon$ and $g'(\epsilon_p)$ is the derivative, i.e., $g'(\epsilon_p)=-M$.

As in the feasible set of $\epsilon$ in (\ref{equ_feasibleset}) we have $W_{\max}+g(\epsilon_p)>0$ in (\ref{approx_derivative2}), computing the root of $U^{\prime a}_c=0$ is equivalent to computing the root of the polynomial part $f(\epsilon) = 0$. The equation $f(\epsilon)=0$ has two real roots as follows:

\begin{equation}\label{root}
\epsilon_L = \frac{W_{\max}}{4M}+\frac{r_1}{2}-\frac{r_2}{2}, \ \ \ \ \epsilon_H = \frac{W_{\max}}{4M}+\frac{r_1}{2}+\frac{r_2}{2},
\end{equation} 
where \\
$$
r_1 = \sqrt{A_1+\frac{A_2}{A_3}+\frac{A_3}{A_4}}, \ \ 
r_2 = \sqrt{2A_1-\frac{A_2}{A_3}-\frac{A_3}{A_4}+\frac{A_5}{4r_1}},
$$
$$
A_1 = \frac{(W_{\max})^2}{4M^2}, \ \ 
A_2 = 2\cdot2^{1/3}W^2_{\max},
$$
$$
A_3 = (216CM^4W_{\max}^2 + 27C^2M^2W_{\max}^4 +  
\sqrt{B_1},\\
$$
$$
B_1 = -864C^3M^6W^6_{\max} +
(108CM^4W_{\max}^2+27C^2M^2W_{\max}^4)^2)^{\frac{1}{3}},
$$
$$
A_4 = 3\cdot2^{1/3}CM^2, \ \ 
A_5 = \frac{32\sigma_W}{CM} + \frac{W_{\max}^3}{M^3}.\\
$$

Next, we show that there is a unique root of $f(\epsilon)=0$ in the feasible set $(0,W_{\max}/M)$ (which is equivalent to (\ref{equ_feasibleset})), while the other one is not in the feasible set. First, we show that  the function $f(\epsilon)$ is continuous,  $f(0) \cdot f(W_{\max}/M) <0$, and $f(W_{\max}/M)\cdot f(+\infty)<0$. This implies that the equation $f_1(\epsilon)=0$ has at least one root in $(0,W_{\max}/M)$ and at least one root in $(W_{\max}/M,+\infty)$. Together with (\ref{root}), we know that the unique root in $ (0,W_{\max}/M)$ is $\epsilon_L$.  $\hfill\square$

\subsection{Proof of Corollary \ref{Cor_EUT_cmp}}\label{sec_B}

The proof consists of two steps. We first obtain the approximated optimal solution under the EUT case, similar to that under the PT case. We then compare the approximated optimal solutions of both cases.

\textbf{Step 1}: We obtain the approximated optimal solution under the EUT case similarly to that under the   PT case.  Recall that when $\lambda=1$ and $\beta=1$, (\ref{pro_valuation}) corresponds to the EUT representation. We denote the privacy level of participation with the EUT representation under the $\epsilon$-differentially private mechanism as $\epsilon_{e}$ (we use this to characterize the feasible set later for the ease of presentation). The privacy cost of participation for the EUT case is given by $g(\epsilon_e)=-M_e\epsilon$ ,where  $M_e=\frac{c}{2}$ based on (\ref{prospect_level2}).

Let
\begin{equation}
f_e(\epsilon) = \left(\frac{W_{\max}+g(\epsilon_e)}{W_{\max}}\right)\left(1+ \frac{kN}{4l}\epsilon^{3}\frac{g'(\epsilon_e)}{W_{\max}}\right)+\epsilon\frac{g'(\epsilon_e)}{W_{\max}} 
\end{equation}
be the polynomial part of the approximated derivative. Here $g'(\epsilon_e)$ is the  derivative, i.e., $g'(\epsilon_e)=-M_e$. Let $\tilde{\epsilon}^*_e$ be the root of $f_e(\epsilon)=0$ in the feasible set $\{\epsilon: -g(\epsilon_e) <W_{\max}\}$, i.e., the approximated optimal solution for the EUT case.

\textbf{Step 2}: We compare the approximated optimal solution $\tilde{\epsilon}^*_e$ for the EUT case with $\tilde{\epsilon}^*$ for the general PT case (excluding the EUT case).  Recall that in the proof of Theorem \ref{Theorem_optimalsolution}, we have $$f(\epsilon) = \left(\frac{W_{\max}+g(\epsilon_p)}{W_{\max}}\right)\left(1+C \frac{M}{W_{\max}}\epsilon^{3}g'(\epsilon_p)\right)+\epsilon\frac{g'(\epsilon_p)}{W_{\max}}$$ and $\tilde{\epsilon}^*$ is the root of $f(\epsilon)=0$. Based on the difference of characteristics between PT modeling and EUT modeling, we have

\begin{equation}
\begin{aligned}
f_e(\tilde{\epsilon}^*_e) & =\left(\frac{W_{\max}+g(\epsilon_{p} )}{W_{\max}}\right)\left(1+ \frac{kN}{4l}\tilde{\epsilon}^{*3}_e\frac{g'(\epsilon_{p})}{W_{\max}}\right)+\tilde{\epsilon}^*_e\frac{g'(\epsilon_{p})}{W_{\max}} \\ &<\left(\frac{W_{\max}+g(\epsilon_{e})}{W_{\max}}\right)\left(1+ \frac{kN}{4l}\tilde{\epsilon}^{*3}_e\frac{g'(\epsilon_{e})}{W_{\max}}\right)+\tilde{\epsilon}^*_e\frac{g'(\epsilon_{e})}{W_{\max}}\\
&=f_e(\tilde{\epsilon}^*_e)=0. 
\end{aligned}
\end{equation}

The inequality holds for the following reasons. First, we have  
\begin{equation}
g(\epsilon_p)<g(\epsilon_e) \Leftrightarrow -\frac{\lambda c }{2}\epsilon^{\beta}<-\frac{c}{2}\epsilon.
\end{equation}
The absolute value of privacy cost under the general Prospect Theory modeling is larger than that under the  EUT modeling. Second, we have 
\begin{equation}
g'(\epsilon_p)<g'(\epsilon_e) \Leftrightarrow -\frac{\lambda c \beta}{2} \epsilon ^{\beta-1} < -\frac{c}{2}.
\end{equation}
The increasing rate of the absolute value of privacy cost under general Prospect Theory modeling is larger than that under EUT modeling.

Since $f_1(0)>0$ and $f_1(\tilde{\epsilon}^*_e)<0$, we have$f_1(0) \cdot f_1(\tilde{\epsilon}^*_e)<0$. So $\tilde{\epsilon}^*$, the root of $f_1(\epsilon)=0$, is in the interval $(0,\tilde{\epsilon}^*_e)$, i.e., $\tilde{\epsilon}^*<\tilde{\epsilon}^*_e$.  This completes the proof.  $\hfill\square$

\subsection{Proof of Corollary \ref{Cor_lambda}}\label{sec_C}
 The proof consists of two steps. We first characterize the approximated optimal solution. We then compare the approximated optimal solutions of both cases.

\textbf{Step 1}:  We characterize the approximated optimal solution through the approximated derivative of the objective function under both cases.  Consider two cases: one with a  higher value of parameter $\lambda_1$ and the other with a lower value of $\lambda_2$. Let 
\begin{equation}
f_{\lambda_1}(\epsilon) = \left(\frac{W_{\max}+g(\epsilon_{\lambda_1})}{W_{\max}}\right)\left(1+ \frac{kN}{4l}\epsilon^{3}\frac{g'(\epsilon_{\lambda_1})}{W_{\max}}\right)+\epsilon\frac{g'(\epsilon_{\lambda_1})}{W_{\max}} 
\end{equation}
be the corresponding polynomial part of the approximated derivative for the case of $\lambda = \lambda_1$. Here the privacy cost is $g(\epsilon_{p_3}) = -\lambda_1c(m+1)\epsilon^\beta/m$. Let 
\begin{equation}
f_{\lambda_2}(\epsilon) = \left(\frac{W_{\max}+g(\epsilon_{\lambda_2})}{W_{\max}}\right)\left(1+ \frac{kN}{4l}\epsilon^{3}\frac{g'(\epsilon_{\lambda_2})}{W_{\max}}\right)+\epsilon\frac{g'(\epsilon_{\lambda_2})}{W_{\max}} 
\end{equation}
be the corresponding polynomial part of the approximated derivative for the case of $\lambda = \lambda_2$. Here the privacy cost is $g(\epsilon_{\lambda_2}) = -\lambda_2c(m+1)\epsilon^\beta/m$. 

Let $\tilde{\epsilon}^*_{\lambda_1}$ be the root of $f_{\lambda_1}(\epsilon)=0$ and $\tilde{\epsilon}^*_{\lambda_2}$ be the root of $f_{\lambda_2}(\epsilon)=0$. Then $\tilde{\epsilon}^*_{\lambda_1}$ and $\tilde{\epsilon}^*_{\lambda_2}$ are the approximated optimal solution of the case of $\lambda = \lambda_1$ and $\lambda = \lambda_2$, respectively.

\textbf{Step 2}: We compare the approximated optimal solutions of both cases through comparing the polynomial part of the approximated   derivatives.  Based on the difference of the PT characteristics between both cases, we have

\begin{equation}
\begin{aligned}
f_{\lambda_1}(\tilde{\epsilon}^*_{\lambda_2}) & =\left(\frac{W_{\max}+g(\epsilon_{\lambda_1} )}{W_{\max}}\right)\left(1+ \frac{kN}{4l}\tilde{\epsilon}^{*3}_{\lambda_2}\frac{g'(\epsilon_{\lambda_1})}{W_{\max}}\right)+\tilde{\epsilon}^*_{\lambda_2}\frac{g'(\epsilon_{\lambda_1})}{W_{\max}} \\ &<\left(\frac{W_{\max}+g(\epsilon_{\lambda_2})}{W_{\max}}\right)\left(1+ \frac{kN}{4l}\tilde{\epsilon}^{*3}_{\lambda_2}\frac{g'(\epsilon_{\lambda_2})}{W_{\max}}\right)+\tilde{\epsilon}^*_{\lambda_2}\frac{g'(\epsilon_{\lambda_2})}{W_{\max}}\\
&=f_{\lambda_2}(\tilde{\epsilon}^*_{\lambda_2})=0. 
\end{aligned}
\end{equation}

The inequality holds due to the following reasons. First, we have 
\begin{equation}
g(\epsilon_{\lambda_1}) < g(\epsilon_{\lambda_2}) \Leftrightarrow -\lambda_1\frac{c(m+1)}{m}\epsilon^\beta< -\lambda_2\frac{c(m+1)}{m}\epsilon^\beta.
\end{equation} Second, we have

\begin{equation}
g'(\epsilon_{\lambda_1}) < g'(\epsilon_{\lambda_2}) \Leftrightarrow -\beta \lambda_1\frac{c(m+1)}{m}\epsilon^{\beta-1}< -\beta\lambda_2\frac{c(m+1)}{m}\epsilon^{\beta-1}.
\end{equation}

Since $f_{\lambda_1}(0)>0$ and $f_{\lambda_1}(\tilde{\epsilon}^*_{\lambda_2})<0$, we have $f_{\lambda_1}(0)\cdot f_{\lambda_1}(\tilde{\epsilon}^*_{\lambda_2})<0$. So  $\tilde{\epsilon}^*_{\lambda_1}$, the root of $f_{\lambda_1}(\epsilon)=0$,  is in the interval $(0,\tilde{\epsilon}^*_{\lambda_2})$, i.e,  $\tilde{\epsilon}^*_{\lambda_1}<\tilde{\epsilon}^*_{\lambda_2}$. Thus, the approximated optimal solution $\tilde{\epsilon}^*$ decreases in the parameter $\lambda$. This completes the proof. $\hfill\square$

\subsection{Proof of Theorem \ref{The_ref}}\label{sec_E}
The proof consists of two steps. We first obtain the approximated optimal solution. We then compare the approximated optimal solutions of both cases.

\textbf{Step 1}: We characterize the approximated optimal solution through the approximated   derivative of the objective function under the case of a positive reference point. The prospect privacy level of participation based on (\ref{prospect_level2}) is as follow (with subscript $pos$ and superscript $p$):
\begin{equation}
\begin{aligned}
&\epsilon^{p}_{pos}=\frac{t}{m}\sum_{i=1}^{t}\left(\epsilon_{ref}-\frac{i}{m}\epsilon\right)^{\beta}-\left(1-\frac{t}{m}\right)\lambda\sum_{i=t+1}^{m}\left(\frac{i}{m}\epsilon-\epsilon_{ref}\right)^{\beta}, \\& \text{for} \ \ t \ \ \text{satisfying} \ \ \frac{t}{m}\epsilon<\epsilon_{ref} \ \  \text{and} \ \ \frac{t+1}{m}\epsilon>\epsilon_{ref}.
\end{aligned}
\end{equation}
The first summation term  corresponds to the gain part, and the second summation term corresponds to the loss part.

The prospect privacy level of non-participation is as follow (with subscript $pos$ and superscript $n$):
\begin{equation}
\epsilon^{n}_{pos} = \epsilon^{\beta}_{ref}.
\end{equation}

The implementation to obtain the approximated   derivative of the objective function is similar to that under the case of a zero reference point. Let $U^{\prime a}_{pos}(\epsilon)$ (formulated below) be the approximated derivative of the objective function and  $f_{pos}(\epsilon)$  be  the polynomial part in $U^{\prime a}_{pos}(\epsilon)$:

\begin{equation}
U^{\prime a}_{pos}(\epsilon)=\frac{4}{\left(\frac{W_{\max}+g(\epsilon^{p}_{pos})-g(\epsilon^{n}_{pos}))}{W_{\max}}\right)^3 \epsilon ^3 N^2}f_{pos}(\epsilon),
\end{equation}
and

\begin{multline}
f_{pos}(\epsilon) =  \left(W_{\max}+\frac{g(\epsilon^{p}_{pos})-g(\epsilon^{n}_{pos})}{W_{\max}}\right)\times\\ \left(1+ \frac{kN}{4l}\epsilon^3\frac{g'(\epsilon^{p}_{pos})-g'(\epsilon^{n}_{pos})}{W_{\max}}\right)- \frac{g'(\epsilon^{n}_{pos})-g'(\epsilon^{p}_{pos})}{W_{\max}}\epsilon.
\end{multline}
Let $\tilde{\epsilon}^*_{pos}$ be the  root of $f_{pos}(\epsilon)=0$ in the feasible set  $\{\epsilon:g(\epsilon^{n}_{pos})-g(\epsilon^{p}_{pos}) <W_{\max}\}$. i.e., the approximated optimal solution for the case of a positive reference point.

\textbf{Step 2}: We compare the approximated optimal solutions of both cases through comparing the polynomial part of the approximated   derivatives.  Recall that $\tilde{\epsilon}^*$ the root of $f(\epsilon)=0$, i.e., the approximated optimal solution under the case of a zero reference point. If 
\begin{equation}
f_{pos}(\tilde{\epsilon}^*)<0,
\end{equation}
since $f_{pos}(0)>0$, then we have $f_{pos}(0)\cdot f_{pos}(\tilde{\epsilon}^*)<0$, i.e., $\tilde{\epsilon}^*_{pos}$ is in the interval $(0,\tilde{\epsilon}^*)$. So we have $\tilde{\epsilon}^*_{pos}<\tilde{\epsilon}^*$. 
If 
\begin{equation}
f_{pos}(\tilde{\epsilon}^*)>0,
\end{equation}
then $\tilde{\epsilon}^*_{pos}$ is outside the interval $(0,\tilde{\epsilon}^*)$. So we have $\tilde{\epsilon}^*<\tilde{\epsilon}^*_{pos}$. If

\begin{equation}
f_{pos}(\tilde{\epsilon}^*)=0,
\end{equation}
then $\tilde{\epsilon}^*$ is also the root of $f_{pos}(\epsilon)=0$. Based on the definition of $\tilde{\epsilon}^*_{pos}$, we have $\tilde{\epsilon}^*=\tilde{\epsilon}^*_{pos}$. Recall that there is only one root in the feasible set in the proof of Theorem \ref{Theorem_optimalsolution}.  This completes the proof. $\hfill\square$

\end{document}